%% file: lecture-notes.tex
\documentclass[11pt,a4paper]{article}
\usepackage{jheppub}
\usepackage{orcidlink}
\usepackage{comment}
\usepackage{xcolor}
\makeatletter
\gdef\@fpheader{}
\makeatother

\usepackage[utf8]{inputenc}
\usepackage{textcomp}
\usepackage{amsmath,amssymb}
\usepackage{tabularx}
\usepackage{booktabs}
\usepackage{amsmath}
\usepackage{amssymb}
\usepackage{enumitem}
\usepackage{braket}
\usepackage{graphicx}
\usepackage{caption}
\usepackage{tabularx}
\usepackage{array}
\newcommand{\be}{\begin{equation}}
\newcommand{\ee}{\end{equation}}

\newcommand{\bk}{\boldsymbol{k}}

\newcommand{\bp}{\boldsymbol{p}}

\usepackage{jheppub}
\usepackage[T1]{fontenc}
\usepackage[capitalize,nameinlink]{cleveref}

\crefname{table}{Table}{Tables}
\Crefname{table}{Table}{Tables}
\crefname{appendix}{Appendix}{Appendices}
\Crefname{appendix}{Appendix}{Appendices}
\crefname{figure}{Fig.}{Figs.}
\Crefname{figure}{Figure}{Figures}

\title{de Sitter Spacetime: Geometry, Causal Structure, and Quantum Fields}

\author[a]{Azadeh Maleknejad\note{Corresponding author: \color{blue}{azadeh.maleknejad@swansea.ac.uk}} \orcidlink{0000-0002-1019-8781}}
\author[b]{and, Ethan Richardson \orcidlink{0009-0004-2499-1897}}

\affiliation[a]{Centre for Quantum Fields and Gravity, Department of Physics,
Swansea University, Singleton Park, Swansea SA2 8PP, United Kingdom}

\affiliation[b]{School of Physics and Astronomy, Cardiff University,
Queen’s Buildings, The Parade, Cardiff CF24 3AA, United Kingdom}

\abstract{de~Sitter spacetime plays a central role in modern cosmology, providing an excellent approximation to both inflation and the present accelerated expansion. It also offers a particularly simple setting in which gravity, spacetime symmetry, and quantum field theory depart qualitatively from their Minkowski-space counterparts.
 These lecture notes introduce de Sitter spacetime, from its geometry and symmetries to quantum fields and perturbations on a fixed \(3+1\)-dimensional de Sitter background. We cover coordinate systems, causal structure, unitary representations, particle production in time-dependent backgrounds, and cosmological perturbations, with emphasis on physical interpretation, applications to cosmology, and how curvature reshapes the notions of mass, vacuum, particles, and late-time observables.
 \newline
 \textit{Based on lectures delivered by AM at the DESY Seminar Workshop, Feb 2026.}
}

\begin{document}

\maketitle
\flushbottom

\newpage
\section{Quantum Fields in de Sitter Space: Context and Motivation}

``This celestial sphere that leaves us in wonder 
is but a lantern in imagination.'' 
\\  ---  Khayyam Neyshabouri\footnote{ Khayyam Neyshabouri, a medieval Persian polymath and poet, united the study of the cosmos with poetic reflections on its vastness, mystery, and the limits of human knowledge.
}  (11th--12th century)
\vskip 0.3cm

Relativistic cosmology began in 1917, when Einstein proposed a static universe \cite{Einstein1917} and de Sitter found a distinct matter-free solution with a positive cosmological constant \cite{deSitter1917}. Friedmann \cite{Friedmann1922} and Lemaître \cite{Lemaitre1927} subsequently uncovered expanding solutions, with Lemaître connecting them to galactic redshifts—a picture supported by Hubble’s 1929 distance–redshift relation \cite{Hubble1929}. These developments culminated in the FLRW framework \cite{Robertson1935,Walker1937}, within which de Sitter spacetime is the maximally symmetric limit of exponential expansion. These developments find their modern realization in cosmic inflation, a brief phase of primordial accelerated expansion that explains key cosmological puzzles and is strongly supported, though not directly observed, by current data \cite{Guth:1980zm,Linde:1981mu,Albrecht:1982wi,Planck:2018jri}. Its quasi-de~Sitter geometry stretches quantum fluctuations to cosmic scales, seeding the observed large-scale structure \cite{Mukhanov:1981xt,Hawking:1982cz,Guth:1982ec}. The 1998 discovery of late-time acceleration revealed a dominant dark-energy component and raised the possibility that the Universe may approach de~Sitter spacetime again in the far future \cite{Riess:1998cb,Perlmutter:1998np}.

\begin{figure}[t]
    \centering
    \includegraphics[width=0.9\linewidth]{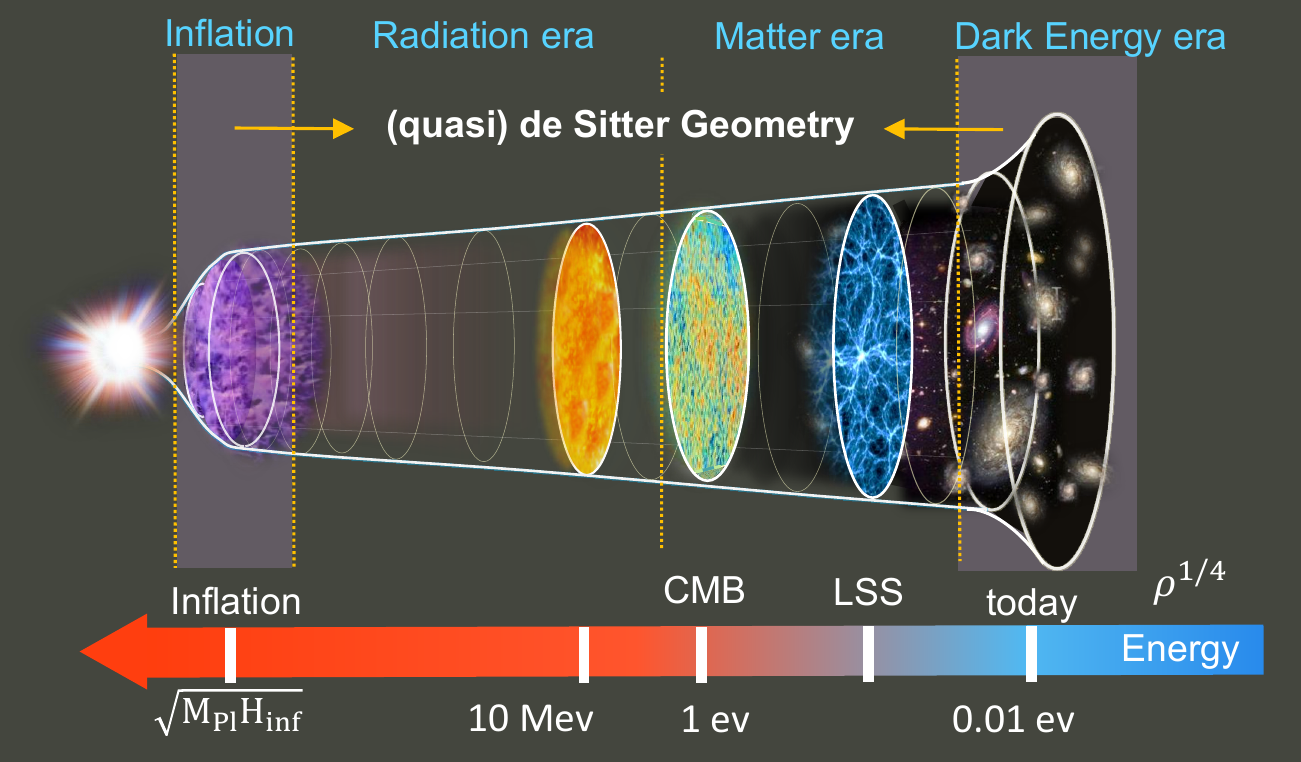}
    \caption{Schematic history of cosmic expansion. Both primordial inflation, preceding the hot Big Bang, and the late-time dark-energy-dominated Universe are described, to a good approximation, by (quasi-)de~Sitter geometry, with the radiation- and matter-dominated epochs connecting these two phases of accelerated expansion. Here $\rho$ denotes the energy density of the Universe during each cosmological era; Big Bang nucleosynthesis (BBN), the cosmic microwave background (CMB),
and large-scale structure (LSS) stand for Big Bang nucleosynthesis, cosmic microwave background, and large-scale structure, respectively.  Fig. adapted with modifications from \cite{Maleknejad:2025clz}.}
    \label{fig:cosmic-history}
\end{figure}

De Sitter space emerges as a unifying framework, capturing both the earliest moments of cosmic evolution and its asymptotic future (see \cref{fig:cosmic-history}).  This dual role makes it essential to understand quantum fields in de Sitter geometry. As such, de Sitter space provides a natural and indispensable setting for exploring the fundamental interplay between gravitation, cosmology, and quantum theory. Despite its apparent simplicity, de Sitter spacetime possesses several features that make it fundamentally different from Minkowski space and render the formulation of quantum field theory conceptually subtle.

Although maximally symmetric, de~Sitter spacetime lacks a globally timelike Killing vector and therefore a universal notion of energy or particles \cite{BirrellDavies1982,ParkerToms2009}. Its cosmological horizon restricts each observer to a finite region and carries the Gibbons--Hawking temperature, revealing a deep connection between geometry and thermodynamics \cite{GibbonsHawking1977}. In quantum field theory, the vacuum is not fixed by geometry alone, while particle content remains observer- and mode-dependent. The expanding geometry adds a further layer: it can produce particles and amplify vacuum fluctuations. In de~Sitter spacetime, sufficiently light modes are stretched beyond the cosmological horizon and freeze, providing the seeds of primordial perturbations. Depending on the field, quantum state, and observable, long-wavelength modes may produce infrared sensitivity or secular terms in perturbative correlation functions; their physical significance and possible implications for backreaction and de~Sitter stability remain subtle and actively debated. De~Sitter spacetime therefore provides the simplest setting in which maximal symmetry coexists with genuinely time-dependent quantum dynamics. These notes develop its geometry and symmetries, the quantization and representation theory of fields, their late-time behavior, and the resulting phenomena of particle production and cosmological perturbations.

The aim of these lecture notes is to build a concise bridge from the geometry
and symmetries of de~Sitter spacetime to the quantum dynamics of fields and
their cosmological signatures, with particular emphasis on physical
interpretation, applications to cosmology, and how spacetime curvature
reshapes the notions of mass, vacuum, particles, and late-time observables.
Throughout these notes, we work in $(3+1)$-dimensional de~Sitter spacetime,
treating its geometry as a fixed semiclassical background and neglecting the
backreaction of quantum fields. The ultraviolet completion of de~Sitter
spacetime remains an open question.%
\footnote{Relevant proposals and constraints include KKLT vacua, the
de~Sitter swampland conjecture, vacuum-lifetime considerations, and the
Festina--Lente bound
\cite{Kachru:2003aw,Obied:2018sgi,Westphal:2007es,Montero:2019ekk}.} Illustrations are used
throughout to make these abstract ideas more tangible and visually intuitive.
All figures and illustrations were prepared by A.M.; those adapted and redrawn
from earlier sources are explicitly identified in their captions.

\paragraph{Notation:} Throughout these notes, $\Gamma^{\rho}{}_{\mu\nu}$ denotes the
Levi--Civita connection, while $R^{\rho}{}_{\sigma\mu\nu}$,
$R_{\mu\nu}$, and $R$ denote the Riemann tensor, Ricci tensor, and
Ricci scalar, respectively. We adopt the metric signature
$(-,+,+,+)$ and the curvature conventions
\begin{align}
R^{\rho}{}_{\sigma\mu\nu}
&=
\partial_{\mu}\Gamma^{\rho}{}_{\nu\sigma}
-\partial_{\nu}\Gamma^{\rho}{}_{\mu\sigma}
+\Gamma^{\rho}{}_{\mu\lambda}
 \Gamma^{\lambda}{}_{\nu\sigma}
-\Gamma^{\rho}{}_{\nu\lambda}
 \Gamma^{\lambda}{}_{\mu\sigma},
\\
R_{\mu\nu}
&=R^{\rho}{}_{\mu\rho\nu},
\qquad
R=g^{\mu\nu}R_{\mu\nu}.
\label{eq:curvature-conventions}
\end{align}
With these conventions, four-dimensional de~Sitter spacetime satisfies
\begin{equation}
R_{\mu\nu\rho\sigma}
=
H^2\left(
g_{\mu\rho}g_{\nu\sigma}
-g_{\mu\sigma}g_{\nu\rho}
\right),
\qquad
R_{\mu\nu}=3H^2g_{\mu\nu},
\qquad
R=12H^2,
\label{eq:ds-curvature}
\end{equation}
where $H^{-1}$ is the de~Sitter curvature radius and
$H^2=\Lambda/3$.

These notes are organized as follows.  \cref{sec:dS-symmetry} introduces the geometry and symmetry group of de~Sitter space and its flat-space limit. \cref{sec:penrose} surveys its principal coordinate systems and causal structure. \cref{sec:dS-qft-casimirs} develops the representation-theoretic description of quantum fields, emphasizing unitarity, allowed masses, the Higuchi bound, and late-time conformal weights. \cref{sec:particle-production-dS} discusses particle creation, cosmological perturbations, and cosmic no-hair. Finally, \cref{sec:conclusions-and-outlook} presents our conclusions, open problems, and suggestions for further reading, while \cref{sec:open-coord} reviews the open FLRW slicing and \cref{sec:bianchi} briefly summarizes the Bianchi family of spatially homogeneous
geometries.

\section{Symmetry Group of de~Sitter Space}
\label{sec:dS-symmetry}

Symmetry plays a central role in quantum field theory, as it determines both the structure of the theory and the classification of its physical states. In flat spacetime, the relevant symmetry group is the Poincar\'e group, whereas in de~Sitter spacetime it is replaced by the de~Sitter group, whose structure reflects the presence of nonzero spacetime curvature. In this section, we review the symmetry group of de~Sitter space and relate it to the Lorentz group in one higher dimension.

\subsection{de~Sitter space as a hyperboloid in 5D ambient space}

\begin{figure}
    \centering
    \includegraphics[width=0.4\linewidth]{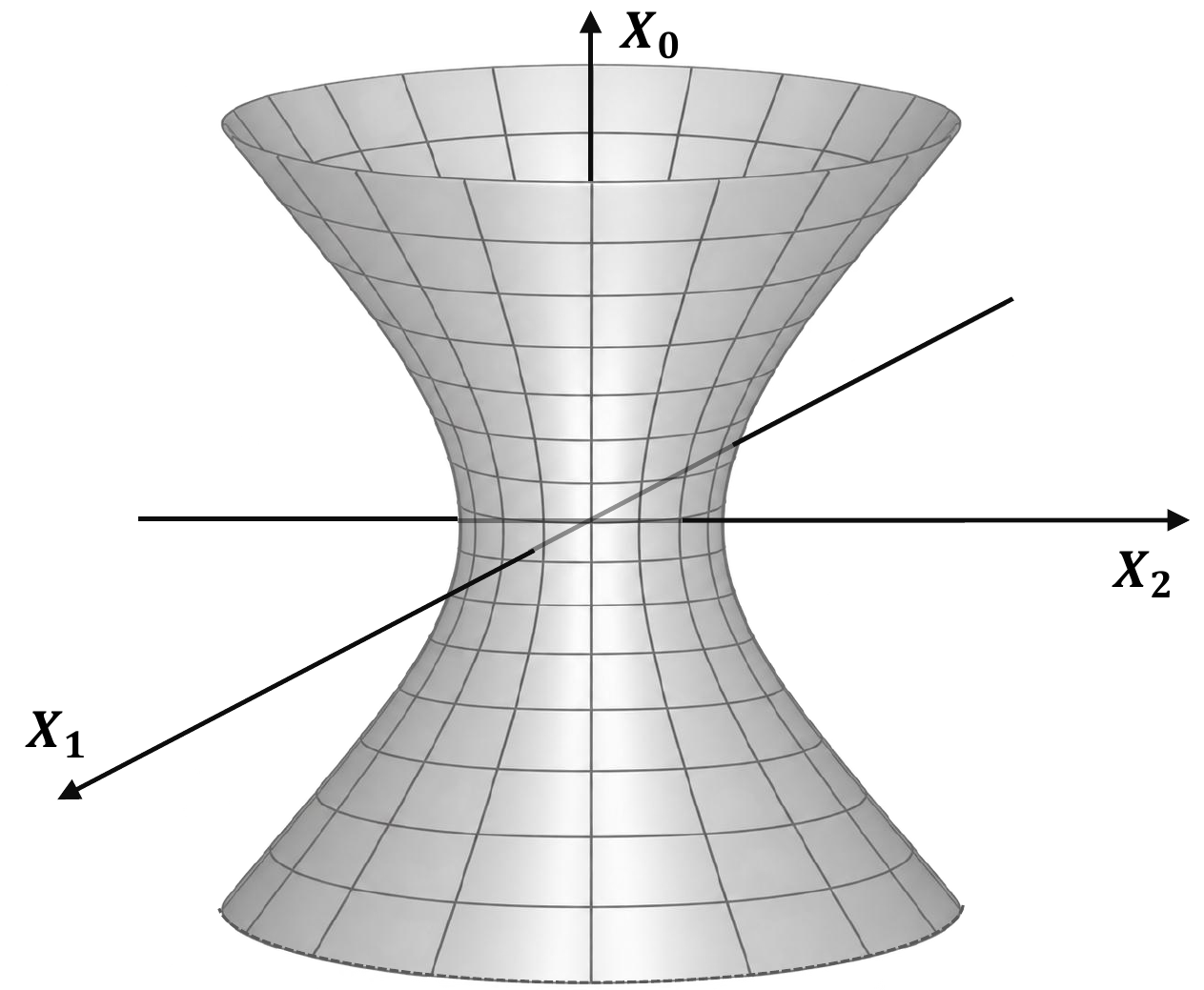}
    \caption{Schematic embedding of four-dimensional de~Sitter space as a hyperboloid in the five-dimensional Minkowski space $\mathbb{R}^{1,4}$. At fixed $X^0$, the spatial section is a three-sphere $S^3$, represented schematically by the bold circle. Two ambient spatial dimensions are suppressed in the illustration. }
    \label{fig:one}
\end{figure}

A particularly useful way to understand de~Sitter space is through its embedding in a higher-dimensional flat spacetime, as illustrated in \cref{fig:one}. Four-dimensional de~Sitter space, $\mathrm{dS}_4$, can be realized as the hyperboloid \cite{Spradlin2001,Kim2002}
\be
\eta_{AB}X^A X^B
=
-(X^0)^2+\sum_{I=1}^{4}(X^I)^2
=
H^{-2},
\label{eq:Hyper}
\ee
embedded in five-dimensional Minkowski space $\mathbb{R}^{1,4}$, with
\be
\eta_{AB}=\mathrm{diag}(-,+,+,+,+),
\qquad
A,B=0,1,2,3,4,
\ee
and line element
\be
\mathrm{d}s^2
=
\eta_{AB}\,\mathrm{d}X^A\,\mathrm{d}X^B
=
-(\mathrm{d}X^0)^2+\sum_{I=1}^{4}(\mathrm{d}X^I)^2.
\ee
Here the de~Sitter radius is $\ell_{\rm dS}=H^{-1}$. We use
\be
X_A\equiv\eta_{AB}X^B,
\qquad
\partial_A\equiv\frac{\partial}{\partial X^A}.
\ee
From this perspective, de~Sitter space is a maximally symmetric spacetime with constant positive curvature, and its symmetries are the transformations that preserve the embedding equation.

Before restricting to the hyperboloid, it is useful to recall the isometries of the ambient spacetime. Five-dimensional Minkowski spacetime is
$\mathbb{M}^{1,4}\equiv\mathbb{R}^{1,4}$, and its connected Poincar\'e group is
\be
\mathrm{ISO}(1,4)
=
\mathrm{SO}(1,4)\ltimes\mathbb{R}^{1,4}.
\label{eq:ambient-Poincare-group}
\ee
Throughout this section, $\mathrm{SO}(1,n)$ denotes the identity component $\mathrm{SO}_0(1,n)$, and $\mathrm{ISO}(1,n)$ denotes the corresponding connected Poincar\'e group. The symbol $\ltimes$ denotes a semidirect product.\footnote{At the Lie-algebra level, let $\mathfrak{g}$ and $\mathfrak{h}$ be Lie algebras and let $
\varphi:\mathfrak{g}\to\mathrm{Der}(\mathfrak{h})$
be a Lie-algebra homomorphism. The semidirect-product Lie algebra $\mathfrak{g}\ltimes_{\varphi}\mathfrak{h}$ is the vector-space direct sum $\mathfrak{g}\oplus\mathfrak{h}$ equipped with the bracket
\begin{equation*}
[(x,u),(y,v)]
=
\left([x,y]_{\mathfrak{g}},
\varphi(x)v-\varphi(y)u+[u,v]_{\mathfrak{h}}\right),
\end{equation*}
for all $x,y\in\mathfrak{g}$ and $u,v\in\mathfrak{h}$.}

The Lie algebra $\mathfrak{iso}(1,4)$ is spanned by ten Lorentz generators,
\be
J_{AB}
=
X_A\partial_B-X_B\partial_A,
\qquad
J_{AB}=-J_{BA},
\label{eq:JAB}
\ee
and five translation generators,
\be
P_A=\partial_A.
\ee
Thus, the five-dimensional Poincar\'e algebra has fifteen generators. Its nonvanishing commutation relations are\footnote{In the equivalent Hermitian convention one defines
$M_{\mu\nu}=iJ_{\mu\nu}$, so that
$M_{\mu\nu}^\dagger=M_{\mu\nu}$. The Lorentz algebra then acquires the usual
explicit factors of $i$ on the right-hand side of the commutation relations.}
\be
[J_{AB},J_{CD}]
=
\eta_{AD}J_{BC}
+\eta_{BC}J_{AD}
-\eta_{AC}J_{BD}
-\eta_{BD}J_{AC},
\ee
\be
[J_{AB},P_C]
=
\eta_{BC}P_A
-\eta_{AC}P_B,
\ee
while the translation generators commute,
\be
[P_A,P_B]=0.
\ee
An element of the connected five-dimensional Poincar\'e group acts on the ambient coordinates $X^A\in\mathbb{R}^{1,4}$ as
\be
X^A
\longrightarrow
{\Lambda^A}_B X^B+a^A,
\qquad
\Lambda\in\mathrm{SO}(1,4),
\qquad
a^A\in\mathbb{R}^{1,4}.
\ee

\subsection{The de~Sitter group as a Lorentz group in 5D}

The de~Sitter hyperboloid must be mapped into itself. Ambient Lorentz transformations preserve $\eta_{AB}X^A X^B$ and therefore preserve the hyperboloid, whereas a generic ambient translation takes a point away from it. The full isometry group and its identity component are
\be
\operatorname{Isom}(\mathrm{dS}_4)=\mathrm{O}(1,4),
\qquad
\operatorname{Isom}_0(\mathrm{dS}_4)=\mathrm{SO}_0(1,4).
\ee
In accordance with the convention stated above, we denote the identity component simply by $\mathrm{SO}(1,4)$. Its Lie algebra $\mathfrak{so}(1,4)$ has ten generators, which are precisely the ambient Lorentz generators in \cref{eq:JAB}. More generally, the connected de~Sitter group in $d$ spacetime dimensions is $\mathrm{SO}(1,d)$, namely the Lorentz group in one higher dimension.

The ambient translation generators $P_A=\partial_A$ do not preserve the hyperboloid and are therefore not isometries of de~Sitter spacetime. Translation-like transformations intrinsic to de~Sitter space are instead generated by the mixed ambient Lorentz generators $J_{\mu d}$. This geometric interpretation will be useful when classifying the unitary irreducible representations of the de~Sitter group.

It is instructive to contrast the de~Sitter group with the symmetry group of flat four-dimensional spacetime. In four-dimensional Minkowski space, the connected symmetry group is the Poincar\'e group,
\be
\mathrm{ISO}(1,3)
=
\mathrm{SO}(1,3)\ltimes\mathbb{R}^{1,3},
\ee
which contains the normal Abelian translation subgroup $\mathbb{R}^{1,3}$ generated by the four-momentum $P_\mu$. This structure gives the usual conserved energy-momentum and leads to the familiar classification of particles by mass and spin \cite{Wigner1939}.

By contrast, the de~Sitter Lie algebra $\mathfrak{so}(1,4)$ is simple and does not decompose as a semidirect product with a translation ideal. In particular, the de~Sitter group contains no normal Abelian subgroup of four mutually commuting spacetime translations. Moreover, de~Sitter space has no globally timelike Killing vector \cite{Spradlin2001}. Consequently, there is no globally defined conserved energy or commuting four-momentum directly analogous to those of Minkowski spacetime, although the ten de~Sitter isometries still give rise to conserved de~Sitter charges. States are instead organized into unitary irreducible representations of $\mathrm{SO}(1,4)$, characterized by spin and the eigenvalues of the Casimir operators; the notion of mass is therefore encoded more subtly than through a Minkowski momentum invariant alone.

We can make contact with the Poincar\'e algebra by decomposing the generators of the de~Sitter group into Lorentz generators and translation-like generators. For $\mathrm{dS}_d$, the generators of $\mathfrak{so}(1,d)$ may be written as
\be
J_{AB}
=
X_A\partial_B-X_B\partial_A,
\qquad
A,B=0,\ldots,d,
\label{eq:dS-generators-general-d}
\ee
where the ambient metric has signature $(-,+,\ldots,+)$. Splitting the indices as
\be
A=(\mu,d),
\qquad
\mu=0,\ldots,d-1,
\ee
we define the translation-like generators
\be
P_\mu\equiv H J_{\mu d}.
\label{eq:P-dS}
\ee
Here $P_\mu$ denotes the rescaled mixed Lorentz generator and should not be confused with the ambient translation operator $P_A=\partial_A$, which is not a de~Sitter isometry. The generators $J_{\mu\nu}$ satisfy the Lorentz algebra $\mathfrak{so}(1,d-1)$ and generate Lorentz transformations in $d$ dimensions. In terms of $J_{\mu\nu}$ and $P_\mu$, the de~Sitter algebra becomes
\begin{align}
[J_{\mu\nu},J_{\rho\sigma}]
&=
\eta_{\mu\sigma}J_{\nu\rho}
+\eta_{\nu\rho}J_{\mu\sigma}
-\eta_{\mu\rho}J_{\nu\sigma}
-\eta_{\nu\sigma}J_{\mu\rho},
\\
[J_{\mu\nu},P_\rho]
&=
\eta_{\nu\rho}P_\mu
-\eta_{\mu\rho}P_\nu,
\\
[P_\mu,P_\nu]
&=
-H^2J_{\mu\nu}.
\label{eq:dS-PP}
\end{align}
The first two commutators have the same form as in the Poincar\'e algebra. The crucial difference is the last relation: the translation-like generators do not commute, and their commutator closes onto the Lorentz generators. Its magnitude is set by
\be
H^2=\ell_{\rm dS}^{-2},
\ee
while the minus sign follows from the mostly-plus convention, for which the additional ambient direction is spacelike and $\eta_{dd}=+1$.

\subsection{Physical interpretation and flat-space limit}

Equation~\eqref{eq:dS-PP} highlights a key conceptual difference between de~Sitter and Minkowski spacetime. In Minkowski space, translations form an Abelian subgroup and are generated by mutually commuting momentum operators. In de~Sitter space, the generators $P_\mu$ do not close among themselves and therefore do not generate a translation subgroup. Nevertheless, de~Sitter space remains homogeneous: its homogeneity is generated by these noncommuting de~Sitter transvections together with the Lorentz generators. Geometrically, $P_\mu$ is proportional to the ambient Lorentz generator $J_{\mu d}$, which mixes the direction $X^\mu$ with the additional embedding direction $X^d$. The factor $H=\ell_{\rm dS}^{-1}$ fixes the normalization of this translation-like generator in terms of the inverse de~Sitter radius. The noncommutativity in \cref{eq:dS-PP} is therefore a direct algebraic manifestation of the constant positive curvature of de~Sitter space. 

The relation between de~Sitter and Minkowski symmetries becomes particularly transparent in the flat-space limit $H\to0$. In this limit, the de~Sitter radius $\ell_{\rm dS}=H^{-1}$ becomes infinite and the hyperboloid becomes locally flat. Holding the rescaled generators $P_\mu=HJ_{\mu d}$ fixed, the commutator $[P_\mu,P_\nu]$ vanishes. At the level of Lie algebras, this defines the In\"on\"u--Wigner contraction \cite{InonuWigner1953}
\be
\mathfrak{so}(1,d)
\longrightarrow
\mathfrak{iso}(1,d-1)
=
\mathfrak{so}(1,d-1)\ltimes\mathbb{R}^{1,d-1}.
\ee
Thus, the de~Sitter algebra can be viewed as a curvature deformation of the Poincar\'e algebra, with the deformation controlled by $H^2=\ell_{\rm dS}^{-2}$. Correspondingly, appropriate families of unitary irreducible representations of the de~Sitter group admit a contraction to representations of the Poincar\'e group characterized by mass and spin; for a discussion of the massive scalar
case, see~\cite{Garidi2003}. The de~Sitter and Poincar\'e groups in the same spacetime dimension have the same number of generators, but their algebraic organization is different: de~Sitter transvections do not commute, whereas Poincar\'e translations do. In the limit $\ell_{\rm dS}\to\infty$, the former contract to the commuting momentum generators of the Poincar\'e algebra. This provides an important consistency check and a useful physical interpretation of the Casimir operators, unitary irreducible representations, and notion of mass in de~Sitter space discussed below.

In summary, de~Sitter space is a maximally symmetric and homogeneous spacetime whose connected isometry group is the Lorentz group in one higher dimension. Although it has the same number of generators as the Poincar\'e group in the corresponding dimension, the absence of a normal Abelian subgroup of commuting spacetime translations leads to important differences in the definitions of energy, momentum, particles, and their spectra.

\medskip

\section{Coordinate Systems and Causal Structure of de~Sitter Space}
\label{sec:penrose}

In this section, we study de~Sitter spacetime from two complementary perspectives. 
First, we review the coordinate systems that are most useful in practice, and then we analyse the global causal structure using the Penrose diagram. 
In \cref{sec:dS-symmetry} we introduced the embedding (ambient) description of de~Sitter space. In this part, we introduce the principal geometric and coordinate descriptions of de~Sitter spacetime. We begin with its Euclidean continuation and FLRW slicings, then discuss the global coordinates covering the complete spacetime, and the cosmological coordinates associated with the Poincar\'e patches. We conclude with the Penrose diagram and the static patch, which make the causal structure and the relation between these coordinate patches manifest. For completeness, we briefly discuss the open FLRW foliation of de~Sitter spacetime in \cref{sec:open-coord}. For pedagogical introductions to de~Sitter geometry, coordinate systems, and causal structure, see Refs.~\cite{Hawking:1973uf,Spradlin2001,Kim2002}. 
For a broader perspective encompassing de~Sitter geometry and holography, with particular emphasis on static-patch and future-boundary viewpoints, see Ref.~\cite{Anninos:2012qw}.

\begin{table}[h]
\centering
\small
\begin{tabularx}{\textwidth}{
|>{\centering\arraybackslash}m{0.10\textwidth}
|m{0.28\textwidth}
|m{0.15\textwidth}
|>{\centering\arraybackslash}m{0.05\textwidth}
|m{0.2745\textwidth}|
}
\hline
\textbf{Slicing}
&
\textbf{Embedding Hyperplane}
&
\textbf{Spatial Slice}
&
$\mathbf{k}$
&
\textbf{Causal Character of $\Sigma_{\perp}$}
\\
\hline
Closed
&
$X_0=\mathrm{const.}$
&
$S^3$
&
$+1$
&
Timelike
\\
\hline
Flat
&
$X_0+X_4=\mathrm{const.}$
&
$\mathbb{R}^3$
&
$0$
&
Null
\\
\hline
Open
&
$X_4=\mathrm{const.}$
&
$H^3$
&
$-1$
&
Spacelike
\\
\hline
\end{tabularx}
\caption{Classification of the three standard spatial slicings of de~Sitter spacetime by the embedding hyperplane $\Sigma_{\perp}$ used to intersect the de~Sitter hyperboloid, the resulting spatial geometry, the normalized spatial curvature parameter $k$, and the causal character of $\Sigma_{\perp}$.}
\label{tab:ds_slicings}
\end{table}

\begin{figure}[b]
  \centering
  \includegraphics[width=0.95\textwidth]{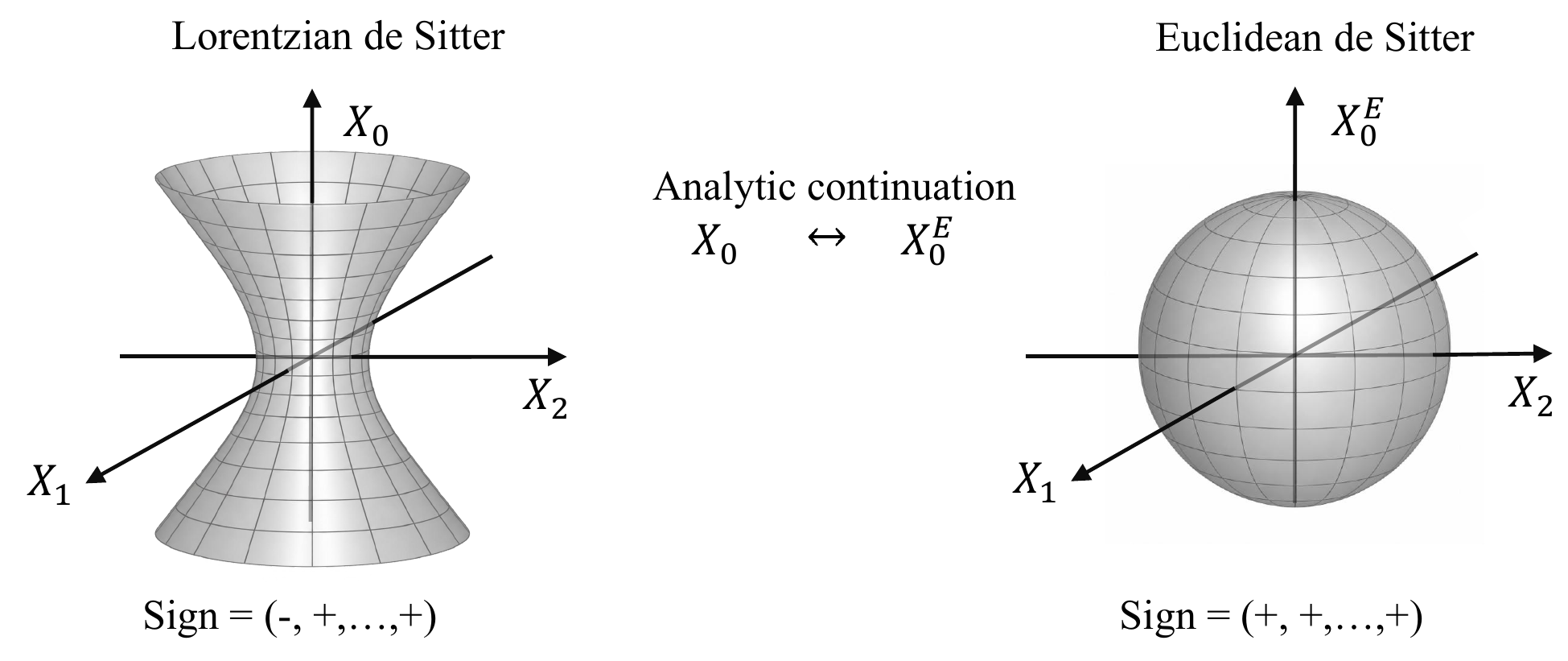}
  \caption{Lorentzian and Euclidean de~Sitter space in the ambient-space picture. 
  Lorentzian $\mathrm{dS}_4$ is realized as a hyperboloid in $\mathbb{R}^{1,4}$, while its Euclidean continuation is the round sphere $S^4\subset\mathbb{R}^5$. 
  The two geometries are related by the analytic continuation $X^0=iX_E^0$, which maps the equator of $S^4$ to the $t=0$ spatial slice of Lorentzian space.}
  \label{fig:lorentz-euclid-dS}
\end{figure}

\subsection{Euclidean de~Sitter space}

Euclidean de~Sitter space is obtained from Lorentzian de~Sitter space by analytic continuation. The relation is particularly transparent in the ambient-space description. Four-dimensional Lorentzian de~Sitter space is represented by the hyperboloid defined in \cref{eq:Hyper}, embedded in five-dimensional Minkowski space. As illustrated in \cref{fig:lorentz-euclid-dS}, analytically continuing the timelike ambient coordinate according to
\be
X^0=iX_E^0
\ee
turns the embedding condition into
\be
\sum_{A=0}^{4}(X_E^A)^2=H^{-2},
\ee
which defines a round four-sphere $S^4$ of radius $H^{-1}$ in Euclidean
space $\mathbb{R}^5$. Thus Euclidean de~Sitter space is simply
$\mathrm{dS}_4^{(E)}\simeq S^4$ \cite{Moschella:2006pkh}.
In particular, the Lorentzian topology $\mathbb{R}\times S^3$ is
replaced by the compact Euclidean topology $S^4$.
The Euclidean section remains maximally symmetric and has the same
positive curvature scale, with
$R=12H^2$, while its isometry group is $\mathrm{SO}(5)$ rather than the
Lorentzian group $\mathrm{SO}(1,4)$.

\begin{figure}[b]
  \centering
\includegraphics[width=0.95\linewidth]{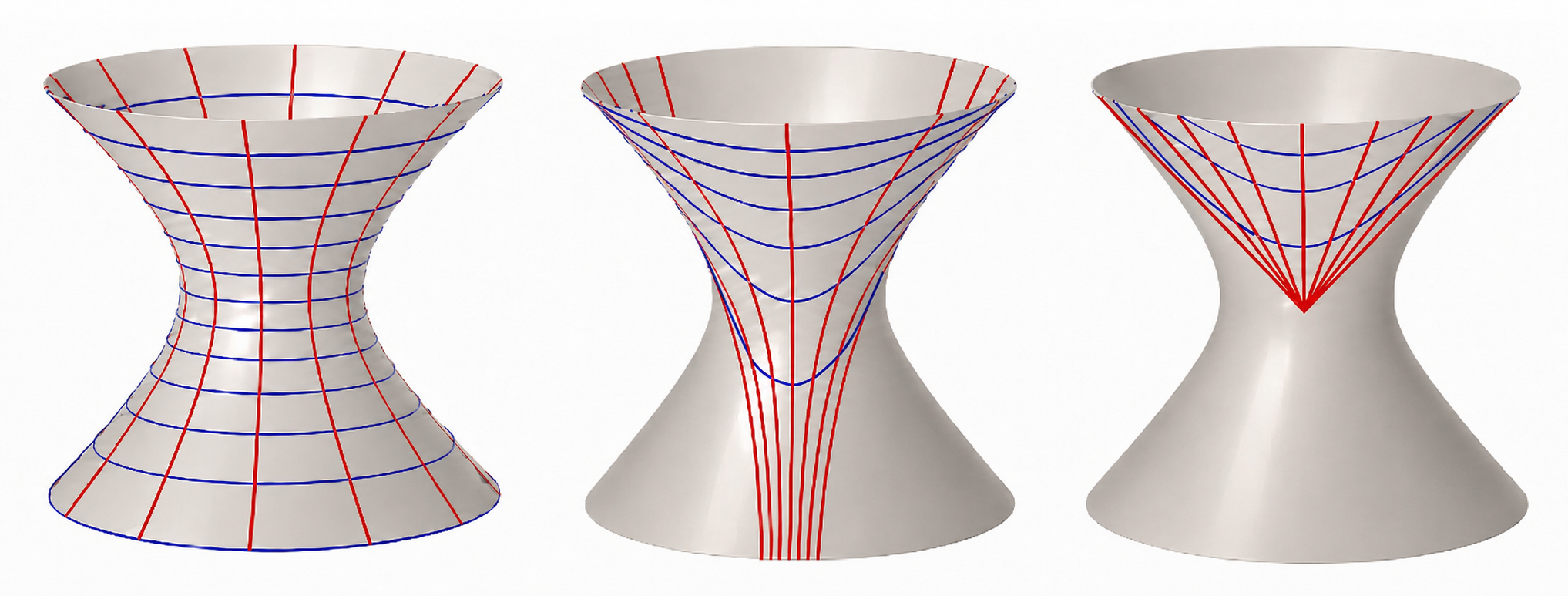}
  \caption{Three different spatial foliations of the same de~Sitter spacetime. Blue curves denote hypersurfaces of constant cosmic time, while red curves denote timelike geodesics. \textbf{Left:} the closed slicing, in which the spatial sections contract to a minimum size and subsequently re-expand. \textbf{Middle:} the flat slicing, describing an exponentially expanding spatially flat patch. \textbf{Right:} the open slicing, with negatively curved, hyperbolic spatial sections. Adapted with modifications from \cite{Hawking:1973uf,Moschella:2006pkh,Gubitosi:2011hgc}.}
  \label{fig:flat-open-close}
\end{figure}

\subsection{FLRW slicings of de~Sitter spacetime}
\label{subsec:dS-slicings}

Different coordinate systems emphasis different aspects of de~Sitter spacetime, and the most convenient choice depends on the physical question at hand. Coordinates adapted to inflationary dynamics, for instance, need not be the most transparent for studying the global causal structure. It is therefore useful to understand the principal coordinate systems used in de~Sitter space, how they are related, and which geometric features each makes manifest. A natural starting point is the embedding of 4D de~Sitter spacetime as the hyperboloid
in 5D Minkowski spacetime (see \cref{eq:Hyper}). Different parameterizations of this same hyperboloid lead to FLRW metrics whose spatial hypersurfaces have positive, zero, or negative intrinsic curvature. These are conventionally referred to as the \emph{closed} (global), \emph{flat} (cosmological), and \emph{open} slicings, respectively, and are illustrated schematically in \cref{fig:flat-open-close}.

More generally, a homogeneous and isotropic spacetime can be written in the Friedmann--Lemaître--Robertson--Walker form
\be
ds^2
=
-dt^2+a^2(t)\,d\Sigma_k^2,
\ee
where $d\Sigma_k^2$ is the metric on a maximally symmetric three-dimensional space. With the conventional normalization $k=+1,0,-1$,
\be
d\Sigma_k^2
=
d\chi^2+S_k^2(\chi)\,d\Omega_2^2,
\ee
in which $S_k(\chi)$ is 
\be
S_k(\chi)
=
\begin{cases}
\sin\chi, & k=+1,\\
\chi, & k=0,\\
\sinh\chi, & k=-1.
\end{cases}
\ee
The corresponding constant-$t$ hypersurfaces are, respectively,
\be
\underbrace{S^3}_{\text{three-sphere}},
\qquad
\underbrace{\mathbb{R}^3}_{\text{Euclidean space}},
\qquad
\underbrace{\mathbb{H}^3}_{\text{hyperbolic space}},
\ee
with intrinsic scalar curvature
\be
{}^{(3)}R
=
\frac{6k}{a^2(t)}.
\ee

For a generic FLRW universe, the matter content naturally selects a preferred cosmological rest frame. For example, a perfect fluid has
\be
T_{\mu\nu}
=
(\rho+p)u_\mu u_\nu+p\,g_{\mu\nu},
\ee
so that, when $\rho+p\neq0$, its four-velocity $u^\mu$ defines a preferred congruence of comoving observers and hence a preferred foliation by homogeneous spatial hypersurfaces. In this setting, the spatial curvature $k/a^2$ has a direct physical meaning.

De~Sitter spacetime is exceptional. The same four-dimensional geometry can be foliated by spatial hypersurfaces of all three possible constant curvatures:
\be
\begin{aligned}
ds^2
&=
-dt^2
+
H^{-2}\cosh^2(Ht)\,d\Omega_3^2,
&& k=+1,
\\[2mm]
ds^2
&=
-dt^2
+
e^{2Ht}\,d\mathbf{x}^2,
&& k=0,
\\[2mm]
ds^2
&=
-dt^2
+
H^{-2}\sinh^2(Ht)\,d\Sigma_{-1}^2,
&& k=-1,
\qquad t>0.
\end{aligned}
\ee
Thus, the spatial slices may be spherical, Euclidean, or hyperbolic, even though the underlying four-dimensional spacetime is the same. In the flat slicing, the arbitrary constant normalization of the scale factor has been absorbed into the dimensionful Cartesian coordinates $\mathbf{x}$, so that $a(t)=e^{Ht}$ is dimensionless in this convention.

This can be seen directly from the Friedmann equations for a spacetime containing only a positive cosmological constant,
\be
\left(\frac{\dot a}{a}\right)^2
+
\frac{k}{a^2}
=
H^2,
\qquad
\frac{\ddot a}{a}
=
H^2,
\qquad
H^2=\frac{\Lambda}{3}.
\ee
The three scale factors
\be
a(t)
=
\begin{cases}
H^{-1}\cosh(Ht), & k=+1,\\
e^{Ht}, & k=0,\\
H^{-1}\sinh(Ht), & k=-1,
\end{cases}
\ee
all satisfy these equations in their respective coordinate domains. With the curvature convention
\be
\begin{aligned}
R^\rho{}_{\sigma\mu\nu}
&=
\partial_\mu\Gamma^\rho_{\nu\sigma}
-
\partial_\nu\Gamma^\rho_{\mu\sigma}
\\
&\quad
+
\Gamma^\rho_{\mu\lambda}\Gamma^\lambda_{\nu\sigma}
-
\Gamma^\rho_{\nu\lambda}\Gamma^\lambda_{\mu\sigma}.
\end{aligned}
\ee
the four-dimensional Riemann tensor is therefore
\be
R_{\mu\nu\rho\sigma}
=
H^2
\left(
g_{\mu\rho}g_{\nu\sigma}
-
g_{\mu\sigma}g_{\nu\rho}
\right),
\ee
and hence
\be
R_{\mu\nu}
=
3H^2g_{\mu\nu},
\qquad
R
=
12H^2.
\label{eq:R-H}
\ee
None of these coordinate-independent quantities depends on $k$. The parameter $k$ therefore characterizes the intrinsic geometry of the chosen spatial slices, rather than an invariant property of the four-dimensional de~Sitter geometry. 

The origin of this freedom can also be understood from the stress tensor of a cosmological constant,
\be
T_{\mu\nu}^{(\Lambda)}
=
-\frac{\Lambda}{8\pi G_{\rm N}}\,g_{\mu\nu}.
\ee
Since this expression contains no preferred timelike four-velocity, the cosmological constant does not select a distinguished cosmological rest frame. Different families of observers can therefore define different notions of simultaneity and, correspondingly, different spacelike foliations of the same de~Sitter manifold.

This multiplicity of descriptions is useful both conceptually and practically. The closed slicing makes the global structure and compact spatial geometry manifest, the flat slicing is particularly well adapted to inflationary cosmology, and the open slicing arises naturally in problems involving bubble nucleation and hyperbolic spatial geometry. Their global properties are also different: the closed coordinates cover the entire de~Sitter hyperboloid, whereas the flat and open coordinates cover only restricted regions. Hence, in de~Sitter spacetime, $k$ labels the choice of FLRW foliation rather than a physically distinct spacetime. The invariant quantity is instead the de~Sitter curvature scale $H^{-1}$, or equivalently the cosmological constant $\Lambda=3H^2$. The main geometric properties of the three standard de~Sitter slicings are summarized in \cref{tab:ds_slicings}.

\subsection{Global coordinates and the complete spacetime}

To make the global causal structure of de~Sitter spacetime manifest---including
its horizons, causal diamonds, and conformal boundaries---we introduce global
coordinates, which cover the entire hyperboloid. They reveal both past and
future conformal infinity, the cosmological horizon associated with each
timelike observer, and the fact that the flat slicing covers only a portion of
the full spacetime.

The closed slicing is obtained by parameterizing the de~Sitter hyperboloid as
\be
\begin{aligned}
X^0 &= H^{-1}\sinh(Ht), \\
X^A &= H^{-1}\cosh(Ht)\,n^A,
\end{aligned}
\ee
where $A=1,\ldots,4$, and
$\sum_{A=1}^{4}(n^A)^2=1$.
Thus, the variables $n^A$ parametrize a unit three-sphere $S^3$. Each
constant-$t$ hypersurface is therefore a three-sphere of physical radius
\be
a(t)=H^{-1}\cosh(Ht).
\ee
The radius decreases from infinity as $t$ approaches zero from the past,
reaches its minimum value
\be
a_{\min}=H^{-1}
\ee
at $t=0$, and then grows again for $t>0$ (see the left panel of
\cref{fig:flat-open-close}). In this foliation, de~Sitter spacetime therefore
appears as a contracting phase followed by an expanding phase. The
corresponding FLRW Hubble parameter is
\be
H_{\rm FLRW}(t)
\equiv
\frac{\dot a}{a}
=
H\tanh(Ht),
\ee
which approaches $-H$ as $t\to-\infty$ and $+H$ as $t\to+\infty$. This slicing is also referred to as the \emph{global slicing}, since
\be
-\infty<t<\infty
\ee
together with the full range of the $S^3$ coordinates covers the entire
de~Sitter hyperboloid. In this sense, the standard global coordinates of
de~Sitter spacetime coincide with the closed FLRW slicing.

\begin{figure}
  \centering
 \includegraphics[width=0.7\linewidth]{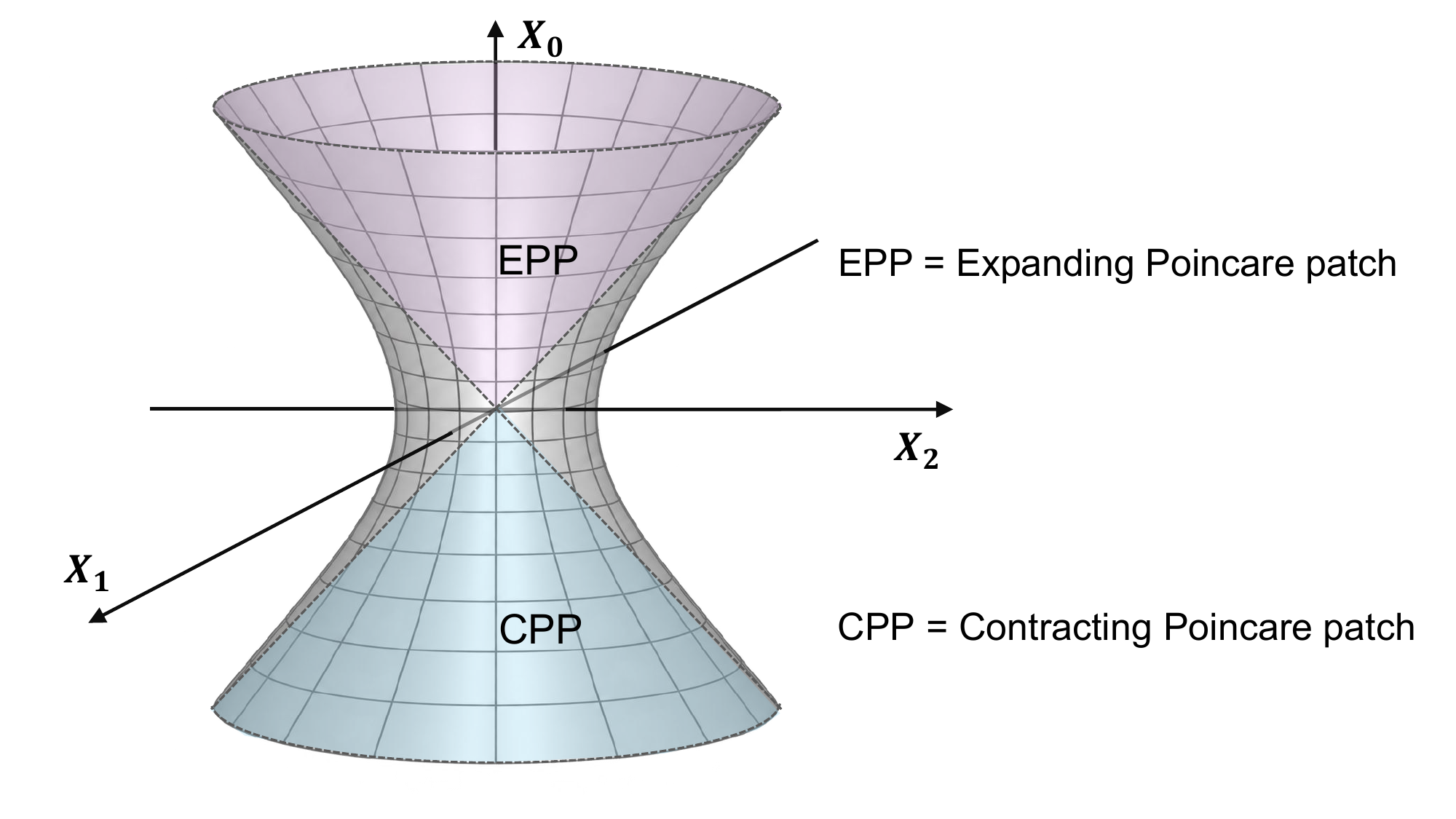}
  \caption{Global view of de~Sitter spacetime, highlighting the two flat FLRW regions: the expanding Poincar\'e patch (EPP) in pink and the contracting Poincar\'e patch (CPP) in blue. These patches provide complementary coordinate descriptions of the expanding and contracting sectors of de~Sitter spacetime.}
  \label{fig:Poincare}
\end{figure}

It is also useful to introduce global conformal time $\tau$ through
\be
\tan\tau=\sinh(Ht),
\qquad
-\frac{\pi}{2}<\tau<\frac{\pi}{2}.
\ee
Since
\be
\cosh(Ht)=\frac{1}{\cos\tau},
\ee
the metric becomes
\be
ds^2
=
\frac{1}{H^2\cos^2\tau}
\left(
-d\tau^2+d\Omega_3^2
\right).
\ee
Writing
\be
d\Omega_3^2
=
d\theta^2+\sin^2\theta\,d\Omega_2^2,
\qquad
0\leq\theta\leq\pi,
\ee
and choosing the orientation
\be
n^i=\sin\theta\,\widehat n^i,
\qquad
n^4=-\cos\theta,
\qquad
\delta_{ij}\widehat n^i\widehat n^j=1,
\qquad
i=1,2,3,
\ee
makes the relation to the flat slicing particularly transparent. In this
orientation,
\be
X^0+X^4
=
\frac{\sin\tau-\cos\theta}{H\cos\tau},
\ee
so the null surface
\be
\tau=\frac{\pi}{2}-\theta
\ee
divides global de~Sitter into the two Poincar\'e patches. The region
\be
\tau>\frac{\pi}{2}-\theta
\ee
is the \emph{expanding Poincar\'e patch} (EPP), corresponding to the usual
expanding flat slicing, whereas
\be
\tau<\frac{\pi}{2}-\theta
\ee
is the \emph{contracting Poincar\'e patch} (CPP), obtained by the
time-reversed flat slicing. As we show in \cref{fig:Poincare}, each Poincar\'e patch therefore covers only one
half of the complete de~Sitter spacetime, while the global coordinates cover
both simultaneously.

\subsection{Cosmological coordinates and Poincar\'e patches}

The coordinate system most commonly used in inflationary cosmology is the
flat, or planar slicing. 
A flat slicing of de~Sitter spacetime is obtained by introducing coordinates
$(t,\mathbf{x})$, with $\mathbf{x}\in\mathbb{R}^3$, through
\be
\begin{aligned}
X^0
&=
H^{-1}\sinh(Ht)
+
\frac{H}{2}e^{Ht}\mathbf{x}^2,
\\
X^i
&=
e^{Ht}x^i,
\\
X^4
&=
H^{-1}\cosh(Ht)
-
\frac{H}{2}e^{Ht}\mathbf{x}^2,
\end{aligned}
\ee
where $i=1,2,3$ and $\mathbf{x}^2\equiv\delta_{ij}x^i x^j$. Each constant-$t$ hypersurface is therefore a copy of Euclidean space
$\mathbb{R}^3$. Its intrinsic curvature vanishes,
\be
{}^{(3)}R=0,
\ee
even though the four-dimensional spacetime is curved. In particular,
  de~Sitter spacetime satisfies \cref{eq:R-H}. 
Thus, the spatial flatness of the constant-$t$ slices should not be
confused with flatness of the full spacetime. In this slicing, the scale factor grows exponentially and the FLRW
Hubble parameter is constant,
\be
H_{\rm FLRW}
\equiv
\frac{\dot a}{a}
=
H.
\ee
This is the form of de~Sitter spacetime most commonly used in
inflationary cosmology. Unlike the closed slicing, planar coordinates cover only part of the de~Sitter hyperboloid. 

The embedding relations imply
\be
X^0+X^4
=
H^{-1}e^{Ht}
>0,
\ee
so this chart covers the region $X^0+X^4>0$, i.e. EPP. Its boundary,
\be
X^0+X^4=0,
\ee
is a null hypersurface called the Poincar\'e horizon. The time-reversed flat slicing covers the complementary region $X^0+X^4<0$, i.e. CPP; see the middle panel of \cref{fig:flat-open-close} and \cref{fig:Poincare}. In the EPP, introducing planar conformal time $\eta$ through
\be
\eta
=
-\frac{1}{H}e^{-Ht},
\qquad
-\infty<\eta<0,
\ee
the scale factor becomes
\be
a(\eta)
=
-\frac{1}{H\eta},
\ee
and the metric takes the conformally flat form
\be
ds^2
=
a^2(\eta)
\left(
-d\eta^2+d\mathbf{x}^2
\right)
=
\frac{1}{H^2\eta^2}
\left(
-d\eta^2+d\mathbf{x}^2
\right).
\label{eq:flat_conformal_metric}
\ee
At fixed $\eta$, the spatial sections are flat and noncompact,
$\Sigma_\eta\simeq\mathbb{R}^3$, and the patch has topology
\be
\mathcal{M}_{\rm EPP}
\simeq
\mathbb{R}\times\mathbb{R}^3.
\ee
This is therefore also called the flat FLRW slicing of de~Sitter space.

Planar coordinates are particularly convenient for inflationary cosmology and quantum field theory. Their spatial homogeneity allows the fields to be decomposed into Fourier modes labelled by the comoving momentum $\mathbf{k}$, while the asymptotic past limit
\be
-k\eta\longrightarrow\infty,
\qquad
k\equiv|\mathbf{k}|,
\ee
provides the standard positive-frequency prescription defining the Bunch--Davies vacuum.

The planar conformal time $\eta$ is related to the global conformal time $\tau$ and the polar coordinate $\theta$ through
\be
-H\eta
=
\frac{\cos\tau}
{\sin\tau-\cos\theta}.
\ee
Since $\cos\tau>0$ throughout the global coordinate range
$-\pi/2<\tau<\pi/2$, the two Poincar\'e patches are
characterised by
\be
\begin{aligned}
\sin\tau-\cos\theta>0
&\quad\Longleftrightarrow\quad
\eta<0
&&
\text{(EPP)},
\\
\sin\tau-\cos\theta<0
&\quad\Longleftrightarrow\quad
\eta>0
&&
\text{(CPP)}.
\end{aligned}
\ee
Their common Poincar\'e horizon is the null surface
\be
\tau
=
\frac{\pi}{2}-\theta.
\ee

The expanding Poincar\'e patch (EPP) is described by
\be
a_{\rm EPP}(\eta)
=
-\frac{1}{H\eta},
\qquad
-\infty<\eta<0.
\ee
Its scale factor grows from $a\to0$ as $\eta\to-\infty$ at the past
Poincar\'e horizon to $a\to\infty$ as $\eta\to0^{-}$ at future conformal
infinity $\mathcal{I}^{+}$. The contracting Poincar\'e patch (CPP) is the
time-reversed branch, for which the planar conformal time is chosen to be
positive:
\be
a_{\rm CPP}(\eta)
=
\frac{1}{H\eta},
\qquad
0<\eta<\infty.
\ee
Its scale factor decreases from $a\to\infty$ as
$\eta\to0^{+}$ at past conformal infinity $\mathcal{I}^{-}$ to $a\to0$
as $\eta\to\infty$ at the future Poincar\'e horizon (see \cref{fig:Poincare}).
Planar coordinates are therefore ideally suited to inflationary
calculations, but not to questions involving the complete causal structure,
for which global coordinates and the Penrose diagram are essential.

\subsection{Penrose diagram of de~Sitter space}
\label{subsec:penrose_diagram}

\begin{figure}[h]
  \centering
\includegraphics[width=0.6\linewidth]{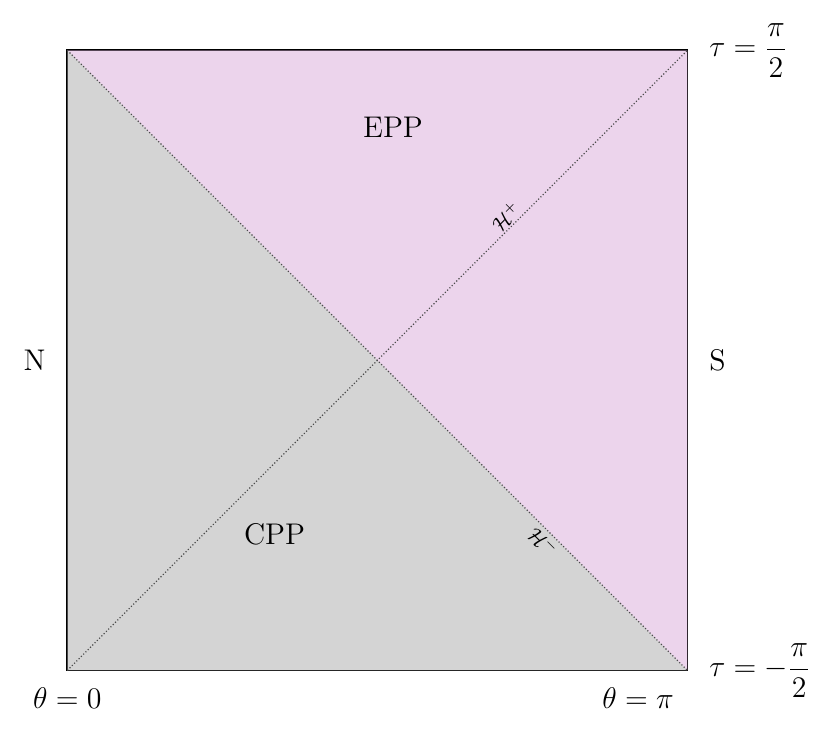}
  \caption{Penrose diagram of global de~Sitter spacetime in conformal coordinates $(\tau,\theta)$, with $\tau\in(-\pi/2,\pi/2)$ and $\theta\in[0,\pi]$. The upper light-orchid region denotes the expanding Poincar\'e patch (EPP), while the lower gray region denotes the contracting Poincar\'e patch (CPP). The dotted null lines indicate the past and future cosmological horizons $\mathcal{H}^{-}$ and $\mathcal{H}^{+}$ of the observer at the south pole. With the orientation adopted here, $\mathcal{H}^{-}$ also coincides with the common Poincar\'e horizon. The symbols $\mathrm{N}$ and $\mathrm{S}$ label the north and south poles of the spatial $S^3$ slices.}
  \label{fig:dS-penrose-2}
\end{figure}

A conformal rescaling preserves null directions and null geodesics as
parametrized curves. The causal structure of de~Sitter spacetime can
therefore be read directly from its conformal metric. In global conformal
coordinates, and with our mostly-plus signature, the metric is
\be
ds^2
=
\frac{1}{H^2\cos^2\tau}
\left(
-d\tau^2+d\theta^2+\sin^2\theta\,d\Omega_2^2
\right),
\qquad
-\frac{\pi}{2}<\tau<\frac{\pi}{2},
\qquad
0\leq\theta\leq\pi.
\ee
Here $\tau$ is the global conformal time and $\theta$ is the polar coordinate
on the spatial three-sphere. After suppressing the $S^2$ angular directions, radial null curves satisfy
\be
ds^2=0
\qquad\Longrightarrow\qquad
d\theta=\pm d\tau.
\ee
Radial light rays therefore propagate at $45^\circ$ in the
$(\tau,\theta)$ plane. Since both $\tau$ and $\theta$ span intervals of length
$\pi$, the global Penrose diagram is a square. The horizontal boundaries
\be
\tau=-\frac{\pi}{2}
\qquad\text{and}\qquad
\tau=+\frac{\pi}{2}
\ee
are, respectively, the spacelike past and future conformal boundaries
$\mathcal{I}^-$ and $\mathcal{I}^+$. The vertical edges at $\theta=0$ and
$\theta=\pi$ are not boundaries of spacetime. They are regular timelike
worldlines corresponding to the north and south poles of the spatial
$S^3$, where the suppressed $S^2$ shrinks smoothly to a point. Every
interior point of the reduced diagram represents an $S^2$; see
\cref{fig:dS-penrose-2}.

\begin{figure}[t]
  \centering
  \includegraphics[width=0.7\linewidth]{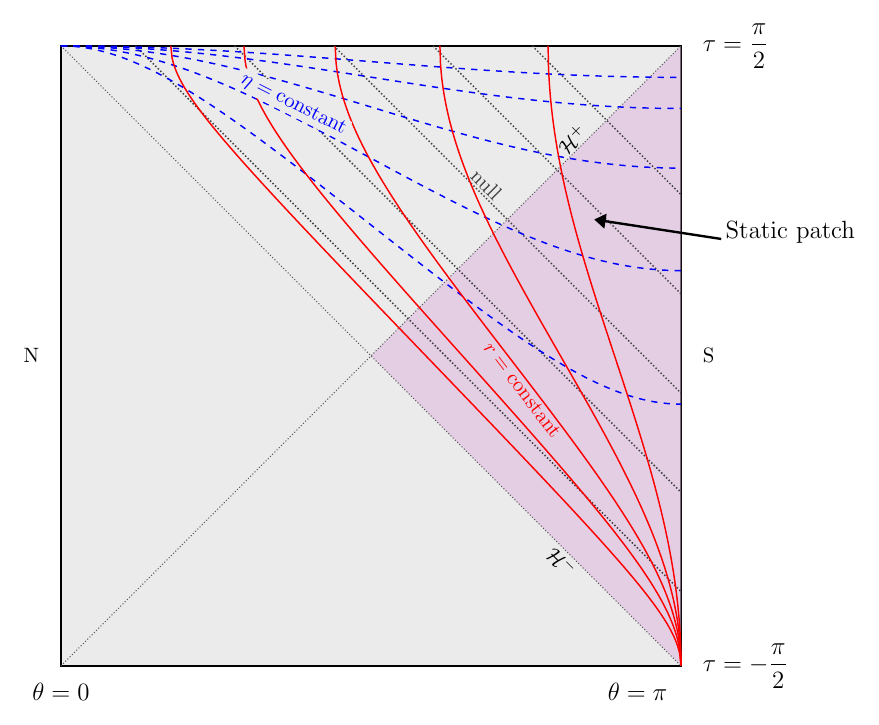}
  \caption{Representation of the global de~Sitter Penrose diagram, analogous to \cref{fig:dS-penrose-2}. The blue dashed curves show slices of constant planar conformal time $\eta$, while the red curves denote comoving worldlines at fixed planar radius $r\equiv|\mathbf{x}|$, and the null rays are shown with dotted gray lines. The light-orchid region highlights the static patch of the observer at the south pole, bounded by the future and past cosmological horizons $\mathcal{H}^{+}$ and $\mathcal{H}^{-}$, shown as dotted null lines. The planar origin $r=0$ lies at $\theta=\pi$ in the EPP and at the antipodal pole $\theta=0$ in the CPP.}
  \label{fig:dS-penrose-3}
\end{figure}

\paragraph{Cosmological horizons and Poincar\'e patches for an observer at $\theta=\pi$:}
Consider the geodesic observer whose worldline lies at the south pole,
$\theta=\pi$. Its past and future endpoints are
\be
p^-_{\rm S}
=
\left(-\frac{\pi}{2},\pi\right)
\in\mathcal{I}^-,
\qquad
p^+_{\rm S}
=
\left(+\frac{\pi}{2},\pi\right)
\in\mathcal{I}^+.
\ee
The observer's past and future cosmological horizons are the two radial null lines
\be
\mathcal{H}^-:
\qquad
\tau
=
\frac{\pi}{2}-\theta,
\qquad\qquad
\mathcal{H}^+:
\qquad
\tau
=
\theta-\frac{\pi}{2}.
\ee
The corresponding causal diamond is therefore
\be
\pi-\theta+|\tau|
<
\frac{\pi}{2}.
\ee
This causal diamond is bounded by $\mathcal{H}^-$ and $\mathcal{H}^+$.
Events outside it cannot both send signals to and receive signals from the
observer over the course of its complete worldline (see \cref{fig:dS-penrose-3}).

The expanding Poincar\'e patch is larger than the static patch. It contains
the static region together with the future triangular region beyond
$\mathcal{H}^+$ in the Penrose diagram. To relate the planar and global descriptions, define
\be
D(\tau,\theta)
\equiv
\sin\tau-\cos\theta,
\ee
where $D(\tau,\theta)$ is a convenient Poincar\'e-patch function whose sign distinguishes the expanding and contracting Poincar\'e patches.
The global and planar conformal times are related by
\be
-H\eta
=
\frac{\cos\tau}
{\sin\tau-\cos\theta}.
\label{eq:eta-tau}
\ee
Since $\cos\tau>0$ in the interior of the global Penrose
diagram, the two planar patches are distinguished by
\be
\begin{aligned}
\mathrm{EPP}:&
\qquad
\eta<0,
\qquad
D(\tau,\theta)>0,
\\
\mathrm{CPP}:&
\qquad
\eta>0,
\qquad
D(\tau,\theta)<0.
\end{aligned}
\ee
The EPP is the causal future of the point
\be
p^-_{\rm S}
=
\left(-\frac{\pi}{2},\pi\right)
\in\mathcal{I}^-,
\ee
whereas the CPP is the causal past of the antipodal point
\be
\left(+\frac{\pi}{2},0\right)
\in\mathcal{I}^+.
\ee
Their common boundary is the null Poincar\'e horizon
\be
D(\tau,\theta)=0
\qquad\Longleftrightarrow\qquad
\sin\tau-\cos\theta=0
\qquad\Longleftrightarrow\qquad
\tau
=
\frac{\pi}{2}-\theta.
\ee
With the orientation adopted here, this null surface coincides with
$\mathcal{H}^-$ of the observer at $\theta=\pi$. In the flat metric given in \cref{eq:flat_conformal_metric}, the scale factor may
be written uniformly as
\be
a(\eta)=\frac{1}{H|\eta|}.
\ee
It grows toward the future in the EPP, where $\eta$ runs from $-\infty$
to $0^-$, and decreases toward the future in the CPP, where $\eta$ runs
from $0^+$ to $+\infty$. Each planar chart covers only one half of the
global de~Sitter manifold and is geodesically incomplete, although the
underlying global spacetime is smooth. The EPP is the patch most commonly
used in inflationary cosmology.

\subsection{The static patch}
\label{subsec:static_patch}

While global coordinates cover the entire de~Sitter manifold, static
coordinates are adapted to the region accessible to a single observer.
For the geodesic observer at the south pole, $\theta=\pi$, this region is
the causal diamond
\be
\mathcal{D}_{\rm S}
=
\left\{
(\tau,\theta):
-\frac{\pi}{2}<\tau<\frac{\pi}{2},
\quad
\pi-\theta<\frac{\pi}{2}-|\tau|
\right\}.
\ee
Equivalently,
\be
\pi-\theta+|\tau|<\frac{\pi}{2}.
\ee
Its null boundaries are
\be
\mathcal{H}^{-}:
\qquad
\tau=\frac{\pi}{2}-\theta,
\qquad\qquad
\mathcal{H}^{+}:
\qquad
\tau=\theta-\frac{\pi}{2},
\ee
which are, respectively, the past and future cosmological horizons of the
observer. The static patch is therefore the maximal region from which the
observer can receive signals and to which signals can be sent during the
observer's complete history. The narrowing of the causal diamond near
$\mathcal{I}^{\pm}$ does not represent a physical contraction of spacetime;
it reflects the finite conformal time available for causal communication.

Unlike the EPP and CPP, which are defined by spatially flat cosmological
slicings, the static patch is observer dependent. With the conventions
introduced above, the static patch centred at $\theta=\pi$ lies entirely
within the EPP, while the antipodal static patch centred at $\theta=0$
lies within the CPP.

Within $\mathcal{D}_{\rm S}$, static coordinates $(t_{\rm s},\rho)$ are
related to the global conformal coordinates by
\be
\rho
=
\frac{\sin\theta}{H\cos\tau},
\qquad
\tanh(Ht_{\rm s})
=
-\frac{\sin\tau}{\cos\theta}.
\ee
The minus sign in the second relation reflects the choice of the south pole
as the center of the static patch; along the observer's worldline
$\theta=\pi$, it gives
$\tanh(Ht_{\rm s})=\sin\tau$, so that $t_{\rm s}$ increases toward the
future together with the global conformal time.

The metric then takes the time-independent form
\be
ds^2
=
-\left(1-H^2\rho^2\right)dt_{\rm s}^2
+
\frac{d\rho^2}{1-H^2\rho^2}
+
\rho^2d\Omega_2^2,
\qquad
0\leq\rho<H^{-1}.
\label{eq:static_patch_metric}
\ee
The observer follows the geodesic worldline $\rho=0$, corresponding to
$\theta=\pi$, while $\rho=H^{-1}$ is the cosmological horizon. The apparent
singularity there is purely coordinate: the spacetime curvature remains
finite.

The vector field
\be
\xi=\partial_{t_{\rm s}}
\ee
is a Killing vector that is timelike inside the static patch and becomes
null at the horizon. It therefore defines a natural conserved energy for
particles and fields within the patch and makes static coordinates the
appropriate framework for studying observer-dependent physics and horizon
thermodynamics.

Although the static-patch metric in \cref{eq:static_patch_metric} resembles the
Schwarzschild metric, their causal interpretations are different. A
black-hole horizon conceals an interior region containing the singularity,
whereas the de~Sitter horizon surrounds the observer and limits access to
regions beyond the observer's cosmological horizon. In particular, the de~Sitter horizon lies
at the finite areal radius $\rho=H^{-1}$; it is neither a curvature
singularity nor future conformal infinity.

In quantum field theory, the Bunch--Davies vacuum restricted to a single
static patch is thermal with respect to the static time $t_{\rm s}$.
The geodesic observer at $\rho=0$ measures the Gibbons--Hawking
temperature \cite{GibbonsHawking1977}
\be
T_{\rm dS}
=
\frac{H}{2\pi}.
\ee
A static observer at fixed $\rho>0$ measures the Tolman local temperature \cite{TolmanEhrenfest1930}
\be
T_{\rm loc}(\rho)
=
\frac{H}{2\pi\sqrt{1-H^2\rho^2}}.
\ee
Semi-classically, the thermal character arises because the global vacuum
contains correlations across the cosmological horizon: restricting
observables to one static patch, or equivalently tracing over inaccessible
degrees of freedom beyond the horizon, produces a mixed thermal state. In
four dimensions, the horizon area and the associated Gibbons--Hawking
entropy are
\be
A_{\rm h}
=
\frac{4\pi}{H^2},
\qquad
S_{\rm dS}
=
\frac{A_{\rm h}}{4G_{\rm N}}
=
\frac{\pi}{G_{\rm N}H^2}.
\ee
Thus, the de~Sitter entropy obeys the same area law as black-hole entropy and is set entirely by the horizon scale $H^{-1}$: a larger $H$ corresponds to a smaller horizon area and hence a smaller entropy.

\medskip

\section{Quantum Fields in de~Sitter Space: Unitarity and Casimir Operators}
\label{sec:dS-qft-casimirs}

Geometry determines how signals propagate through spacetime; symmetry determines how quantum states can be organized. In the preceding sections, we described four-dimensional de~Sitter space, $\mathrm{dS}_4$, as a maximally symmetric hyperboloid with isometry group $\mathrm{SO}(1,4)$ and then explored its global causal structure through its coordinate patches and Penrose diagram. We now turn to the quantum theory and ask a more subtle question: what does it mean to speak of a particle, or more generally a quantum field, in de~Sitter spacetime?

In Minkowski spacetime, one-particle states are classified by the unitary irreducible representations of the Poincar\'e group. The four commuting translation generators provide a conserved four-momentum, while the Casimir operators identify the mass and spin. In de~Sitter spacetime, the corresponding classification is based on the unitary irreducible representations of $\mathrm{SO}(1,4)$. Because de~Sitter space has no globally timelike Killing vector, it admits no globally preferred notion of positive frequency, particle number, or energy analogous to the one defined by time translations in Minkowski spacetime. Nevertheless, the Casimir operators of $\mathrm{SO}(1,4)$ provide invariant labels for free quantum fields and reveal how mass, spin, gauge symmetry, and unitarity are intertwined by the curvature scale $H$.

\subsection{One-particle representations: from Poincar\'e to de~Sitter}
\label{subsec:one-particle-representations}


It is useful to begin with the familiar flat-space construction. The Poincar\'e algebra contains the Lorentz generators $J_{\mu\nu}$ together with four mutually commuting translations. On a one-particle state, the latter are characterized by the physical four-momentum $p^\mu$. The quadratic Casimir may therefore be written as
\be
\mathcal{C}_{\rm P}^{(2)}
\equiv
-p_\mu p^\mu.
\ee
For a massive representation,
\be
\mathcal{C}_{\rm P}^{(2)}=m^2. 
\ee
The spin information is encoded in the Pauli--Lubanski vector,
\be
W_\mu
\equiv
-\frac{i}{2}
\varepsilon_{\mu\nu\rho\sigma}
p^\nu J^{\rho\sigma},
\qquad
\varepsilon_{0123}=+1.
\ee
The factor $-i$ follows from our anti-Hermitian convention for the Lorentz generators. The corresponding quartic Casimir is
\be
\mathcal{C}_{\rm P}^{(4)}
\equiv
W_\mu W^\mu.
\ee
For a massive spin-$s$ representation,
\be
\mathcal{C}_{\rm P}^{(4)}
=
m^2 s(s+1).
\ee
Thus a massive one-particle representation is labelled by its mass $m$ and spin $s$ and contains $2s+1$ polarization states.

The massless case is qualitatively different because no rest frame exists. For the ordinary finite-helicity representations,
\be
p_\mu p^\mu=0,
\qquad
W_\mu W^\mu=0,
\ee
so these two Casimirs alone do not distinguish particles of different helicity. Choosing a standard null momentum $k^\mu$, the subgroup of the Lorentz group that leaves it invariant is
\be
G_k
=
\mathrm{ISO}(2)
=
\mathrm{SO}(2)\ltimes\mathbb{R}^2.
\ee
For finite-helicity representations, the translation-like subgroup $\mathbb{R}^2$ acts trivially, and the physical little-group action reduces to
\be
\mathrm{ISO}(2)
\longrightarrow
\mathrm{SO}(2)
\simeq
\mathrm{U}(1).
\ee
Its one-dimensional irreducible representations are labelled by the helicity $\lambda$. Equivalently,
\be
W^\mu
\lvert\boldsymbol{p},\lambda\rangle
=
\lambda p^\mu
\lvert\boldsymbol{p},\lambda\rangle.
\ee
A massless scalar has $\lambda=0$, while a Weyl fermion carries one of the two helicities $\lambda=+\tfrac12$ or $\lambda=-\tfrac12$, depending on its chirality. A parity-invariant photon or graviton field combines the two representations with helicities $\lambda=\pm1$ or $\lambda=\pm2$, respectively. This flat-space example captures the central representation-theoretic idea: an elementary one-particle sector is an irreducible unitary representation of the spacetime isometry group. The Casimir operators provide invariant labels for massive representations, while in the massless case the helicity is specified by the representation of the little group. For more details see for instance  \cite{Wigner1939,WeinbergQFT1,SchwartzQFT}.

\subsubsection*{De~Sitter Casimir operators}

In $dS_4$, the Poincar\'e group is replaced by the de~Sitter isometry group
$\mathrm{SO}(1,4)$. Its ten generators $J_{AB}$, with
$A,B=0,\ldots,4$, were introduced in \eqref{eq:JAB}. For comparison with the flat-space mass, it is convenient to define the
dimensionful quadratic Casimir as
\be
\mathcal{C}_{\rm dS}^{(2)}
\equiv
\frac{H^2}{2}
J_{AB}J^{AB},
\label{eq:dS-quadratic-Casimir}
\ee
where $H$ is the de~Sitter Hubble scale. The factor $H^2$ gives
$\mathcal{C}_{\rm dS}^{(2)}$ mass dimension two.

The connection with the Poincar\'e Casimir becomes manifest in the flat-space
limit $H\rightarrow0$. In this limit, the generators $J_{\mu\nu}$ remain the
Lorentz generators, while the generators mixing the four-dimensional
directions with the fifth embedding direction give the physical momentum,
\be
p_\mu
=
iH J_{4\mu}.
\ee

For a non-scalar bosonic representation of spin $s\geq1$, it is
conventional to define the mass parameter $m$ such that it reduces to the
ordinary particle mass in the flat-space limit, while the strictly massless
theory corresponds to $m=0$. With this convention, the quadratic Casimir has
eigenvalue
\be
\mathcal{C}_{\rm dS}^{(2)}
=
m^2-2(s-1)(s+1)H^2.
\label{eq:ds_casimir_2}
\ee
For spin two, for example,
\be
\mathcal{C}_{\rm dS}^{(2)}
=
m^2-6H^2.
\ee
This should not be confused with the Higuchi combination $m^2-2H^2$,
which instead controls the norm of the helicity-zero mode.

A second independent Casimir is constructed from the five-dimensional
Pauli--Lubanski vector,
\be
W_A
\equiv
-\frac{i}{8}
\varepsilon_{ABCDE}
J^{BC}J^{DE},
\qquad
\varepsilon_{01234}=+1,
\ee
through
\be
\mathcal{C}_{\rm dS}^{(4)}
\equiv
-H^2 W_AW^A.
\label{eq:ds_casimir_4}
\ee
In the same convention, its eigenvalue is
\be
\mathcal{C}_{\rm dS}^{(4)}
=
s(s+1)
\left[
m^2-s(s-1)H^2
\right].
\ee
Thus, the two Casimirs encode the mass and spin labels of a de~Sitter
representation. In the contraction limit $H\to0$,
\be
\mathcal{C}_{\rm dS}^{(2)}
\longrightarrow
-p_\mu p^\mu
=
m^2,
\qquad
\mathcal{C}_{\rm dS}^{(4)}
\longrightarrow
W_\mu W^\mu
=
m^2s(s+1),
\ee
in agreement with the corresponding Poincar\'e Casimirs for the mostly-plus signature. Unlike in flat
spacetime, however, the quadratic Casimir is not simply the mass squared:
curvature couples to the spin degrees of freedom and shifts its eigenvalue
by an amount of order $H^2$.

\subsection{Unitary representations and allowed masses}
\label{subsec:dS-unitary-series}

Unitarity imposes nontrivial restrictions on the allowed relation between
mass and spin. For a generic massive integer-spin field with $s\geq1$, all
$2s+1$ helicities propagate, and canonical quantization requires each
physical helicity state to have positive norm,
\be
\langle s,\lambda,\bk\,|\,s,\lambda,\bk\rangle>0,
\qquad
\lambda=-s,\ldots,s .
\label{eq:positive-helicity-norm}
\ee
The first obstruction encountered when lowering the mass from the ordinary
massive branch arises in the longitudinal helicity-zero sector. On this
branch its norm contains the factor
\be
\langle s,0,\bk\,|\,s,0,\bk\rangle
=
\mathcal N_s\left(\frac{m^2}{H^2}\right)
\left[m^2-s(s-1)H^2\right],
\qquad
\mathcal N_s>0
\quad\text{for}\quad
m^2>s(s-1)H^2 .
\label{eq:helicity-zero-norm}
\ee
Hence a generic massive representation, for which all helicities are
present, is unitary only if
\be
m^2>s(s-1)H^2 .
\label{eq:general_higuchi_bound}
\ee
This is the strict form of the Higuchi bound appropriate to the ordinary massive branch~\cite{Higuchi:1987cm,DeserWaldron2001}. At the endpoint,
$m^2=s(s-1)H^2$, the helicity-zero state becomes null rather than
negative-norm: an additional gauge symmetry appears and removes it from
the physical spectrum. In the convention used below, this is the $t=0$
partially massless point
\cite{Higuchi:1987cm,Higuchi:1986wu,Deser:1983mm,Deser:2001pe}.
At generic masses below this point the representation is non-unitary,
although the pattern of negative-norm helicities changes as the successive
partially massless values are crossed.

The same structure follows directly from the unitary representation theory
of $SO(1,4)$. For a pedagogical treatment from the perspective of de Sitter
QFT and conformal field theory, see \cite{Sun:2021thf}. Let
$\mathcal F_{\Delta,s}$ denote the symmetric traceless spin-$s$
representation space of weight $\Delta$, with shadow weight
\be
\bar\Delta\equiv3-\Delta .
\ee
For $s\geq1$, equating the quadratic Casimir with its bulk expression gives
\be
\frac{m^2}{H^2}
=
(\Delta+s-2)(s+1-\Delta) =
 \left(s-\frac12\right)^2 
-\left(\Delta-\frac32\right)^2 .
\label{eq:mass-dimension-spin}
\ee
For complex $\Delta$ of the form
\be
\Delta=\frac32+i\mu_s,
\qquad
\mu_s \in\mathbb R ,
\ee
the $SO(1,4)$-invariant inner product is the standard $L^2$ product and is
positive definite. These representations form the principal series,
\be
\frac{m^2}{H^2}
\geq
\left(s-\frac12\right)^2,
\qquad
\text{(principal series)} .
\label{eq:principal-series}
\ee

For real $\Delta$, the appropriate invariant norm instead involves the
spin-$s$ shadow operator,
\be
(\psi,\psi)_{\cal C}
=
\int d^3x_1,d^3x_2,
\psi^{*}_{i_1\cdots i_s}(x_1)
S^+_{\Delta,s}(x_{12})^{i_1\cdots i_s}{}_{j_1\cdots j_s}
\psi^{j_1\cdots j_s}(x_2).
\label{eq:complementary-inner-product}
\ee
In momentum space,
\be
S^+_{\Delta,s}(\bp)
=
p^{2\bar\Delta-3}
\sum_{\ell=0}^{s}
\kappa_{s\ell}(\Delta),
\Pi^{s\ell}(\hat{\bp}),
\ee
where
\be
\kappa_{s\ell}(\Delta)
=
\frac{(\Delta+\ell-1)_{s-\ell}}
{(\bar\Delta+\ell-1)_{s-\ell}} .
\ee
Since the projectors $\Pi^{s\ell}$ resolve the different helicity sectors,
unitarity requires their coefficients to have a common sign. Since
$\kappa_{ss}=1$, they must in fact all be positive. Equivalently,
\be
\frac{\kappa_{s,\ell+1}}
{\kappa_{s,\ell}}
=
\frac{\bar\Delta+\ell-1}
{\Delta+\ell-1}
> 0,
 \qquad
 \ell=0,\ldots,s-1.
 \label{eq:spinning-complementary-positivity}
 \ee
 For $s\geq1$ in $dS_4$, these inequalities give
\be
 1<\Delta<2 .
 \ee
 Writing
 \be \Delta=\frac32+\nu_s, \qquad
 |\nu_s|<\frac12 ,
 \ee
 \cref{eq:mass-dimension-spin} then gives
 \be
 s(s-1)
 <
 \frac{m^2}{H^2}
 <
 \left(s-\frac12\right)^2 ,
 \qquad
\text{(complementary series)} .
\label{eq:complementary-series}
\ee
 Thus the principal and complementary series both have positive,
 nondegenerate norms. Their common boundary at
 $m^2/H^2=(s-\tfrac12)^2$ separates the real and complex branches of
 $\Delta$ and does not signal a loss of unitarity.

A qualitatively different phenomenon occurs at the isolated values
\be
\Delta=1-t,
\qquad
\bar\Delta=2+t,
\qquad
t=0,\ldots,s-1 .
\label{eq:discrete-delta}
\ee
Here we have chosen the $\Delta_-=1-t$ realization; equivalently, the
shadow representation has weight $\Delta_+=2+t$. At these points,
\be
\kappa_{s\ell}=0,
\qquad
\ell=0,\ldots,t ,
\ee
whereas the sectors with $\ell\geq t+1$ retain nonzero positive norm.
The invariant pairing on $\mathcal F_{1-t,s}$ is therefore degenerate.
Its null states form an invariant subspace which is quotiented out to
obtain the physical unitary representation. In the bulk, this shortening
corresponds precisely to the emergence of a gauge symmetry removing the
helicities
\be
0,\pm1,\ldots,\pm t .
\ee
Using \cref{eq:mass-dimension-spin}, the corresponding masses are
\be
\frac{m^2}{H^2}
=
s(s-1)-t(t+1),
\qquad
t=0,\ldots,s-1,
\qquad
\text{(discrete series)} .
\label{eq:partially-massless-masses}
\ee
For $0\leq t\leq s-2$ these are partially massless fields, whereas
$t=s-1$ gives the strictly massless representation, with only the
helicities $\pm s$.\footnote{In the more general representation-theory
terminology these shortened spinning representations belong to the
exceptional series of $SO(1,d+1)$. In $dS_4$, the two parity-related
$SO(1,4)$ irreducible components belong to the discrete series, and we use
the customary $dS_4$ terminology ``discrete series'' throughout.} 

\begin{figure}[h]
\centering
\includegraphics[width=0.95\linewidth]{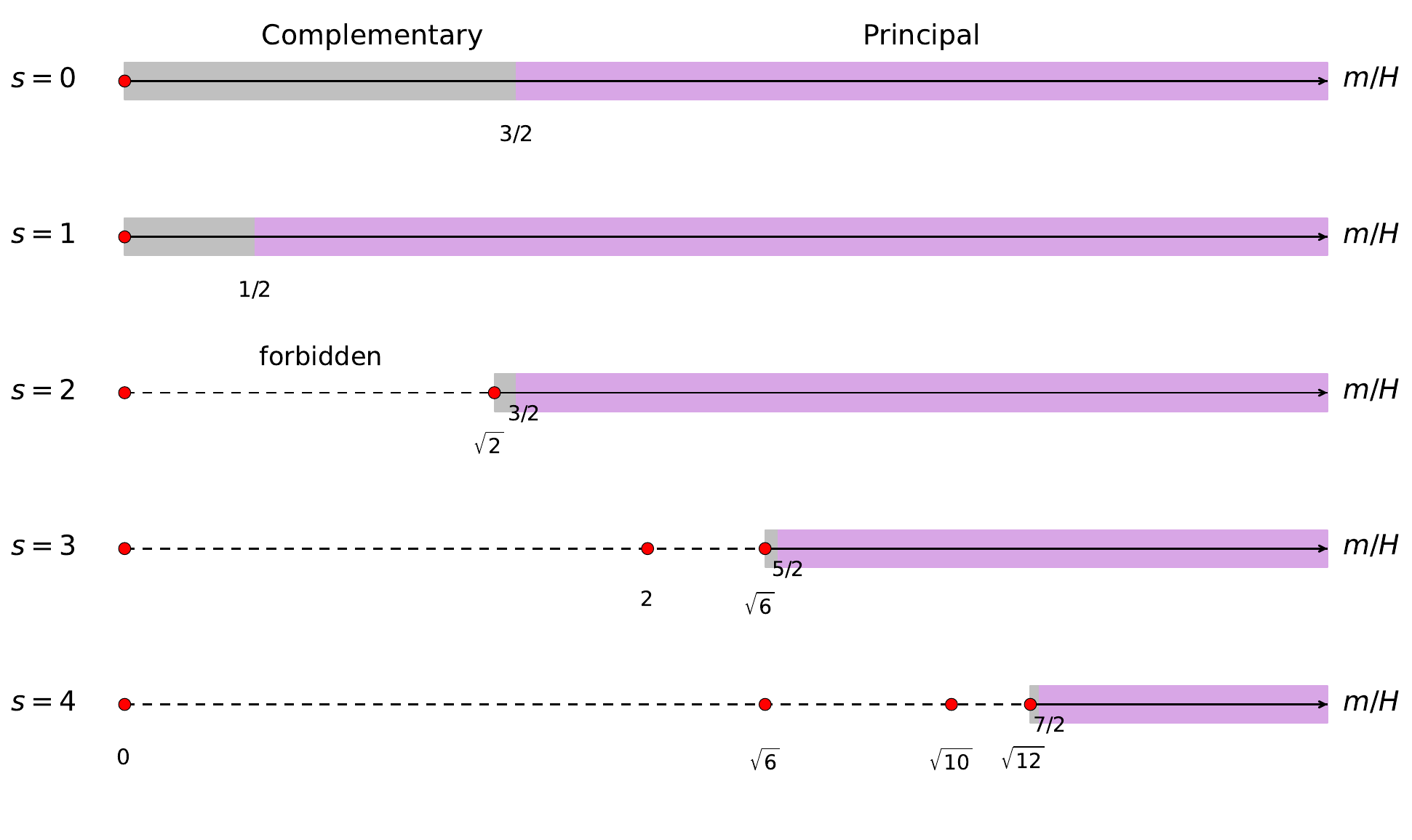}
\caption{Schematic unitary mass spectrum of integer-spin fields in
$dS_4$, shown as a function of $m/H$ for $s=0,\ldots,4$. The silver
and orchid regions denote the complementary and principal series,
respectively, while the dashed intervals denote generic non-unitary
(forbidden) masses. For $s\geq1$, the red points indicate the isolated
unitary gauge representations: the strictly massless points at $m=0$
and, for $s\geq2$, the partially massless points. The scalar
$s=0$, $m=0$ endpoint is special because of its constant zero mode and
should be understood with the caveat discussed in the text.}
\label{fig:unitary-mass-spectrum}
\end{figure}

For a minimally coupled scalar
satisfying
\be
(\Box-M^2)\phi=0 ,
\ee
the mass--dimension relation is instead
\be
\frac{M^2}{H^2}
=
\Delta(3-\Delta).
\label{eq:scalar-mass-dimension}
\ee
For nonnegative Klein--Gordon mass, the scalar principal series has
$M^2/H^2\geq9/4$, while the complementary series has
$0<M^2/H^2<9/4$. The massless minimally coupled endpoint $M=0$ is
special because of its constant zero mode and associated infrared
subtleties, and should not be identified naively with the ordinary
massless spinning representations discussed above. As a simple example, for spin two the generic massive branch obeys
$m^2>2H^2$. The point $m^2=2H^2$ is the $t=0$ partially massless
representation, the point $m^2=0$ is strictly massless, and generic
masses in the interval $0<m^2<2H^2$ are non-unitary. The resulting structure is illustrated schematically in
\cref{fig:unitary-mass-spectrum}, which displays the continuous
complementary and principal branches together with the isolated massless
and partially massless points that remain unitary within the otherwise
forbidden low-mass region.

\subsection{Spin two and the Higuchi bound}
\label{subsec:spin-two-Higuchi}

In \cref{subsec:dS-unitary-series}, we discussed the unitary representations of the de~Sitter group and the restrictions that unitarity places on the mass of a field of given spin. It is useful to pause at this point and examine in more detail the first nontrivial spinning example, namely $s=2$, for which the abstract representation-theoretic bound can be seen directly at the level of the field theory. This is the celebrated \emph{Higuchi bound}.

The linear theory of a massive spin-two field goes back to Fierz and Pauli~\cite{FierzPauli1939}, who identified the special quadratic mass combination required to propagate the correct five degrees of freedom. Nearly five decades later, Higuchi showed that de~Sitter curvature imposes an additional restriction: in $\mathrm{dS}_4$ the range
\be
0<m^2<2H^2=\frac{2}{3}\Lambda
\ee
contains a negative-norm helicity-zero state and is therefore non-unitary~\cite{Higuchi1987}. Curvature-dependent gauge invariance of higher-spin fields had already been studied in related contexts~\cite{DeserNepomechie1984}; the special endpoint $m^2=2H^2$ was later systematically understood as the simplest example of a \emph{partially massless} theory~\cite{DeserWaldron2001,DeserWaldron2001b}.

Consider a symmetric spin-two fluctuation $h_{\mu\nu}$ on a fixed de~Sitter background $\bar g_{\mu\nu}$,
\be
g_{\mu\nu}
=
\bar g_{\mu\nu}+h_{\mu\nu},
\qquad
\bar R_{\mu\nu}
=
3H^2\bar g_{\mu\nu}.
\ee
At quadratic order, the ghost-free mass term is the Fierz--Pauli combination, and the action may be written as~\cite{FierzPauli1939}
\be
S_{\rm FP}^{(2)}
=
\frac12
\int d^4x\sqrt{-\bar g}\,
h^{\mu\nu}
\mathcal{E}_{\mu\nu}{}^{\rho\sigma}
h_{\rho\sigma}
-
\frac{m^2}{4}
\int d^4x\sqrt{-\bar g}\,
\left(
h_{\mu\nu}h^{\mu\nu}-h^2
\right),
\label{eq:dS-Fierz-Pauli-action}
\ee
where $h\equiv \bar g^{\mu\nu}h_{\mu\nu}$ and $\mathcal{E}_{\mu\nu}{}^{\rho\sigma}$ denotes the linearized Einstein operator, including the de~Sitter curvature terms. Its normalization is chosen such that the $m=0$ theory is invariant under linearized diffeomorphisms. For generic masses, i.e.
\be
m^2\neq0,
\qquad
m^2\neq2H^2,
\ee
the equations of motion imply the four vector constraints and one scalar constraint,
\be
\bar\nabla^\mu h_{\mu\nu}=0,
\qquad
h=0.
\ee
These five constraints reduce the ten components of a symmetric tensor to the five physical helicities
\be
\lambda=\pm2,\ \pm1,\ 0
\ee
of a massive spin-two particle.

To see explicitly where the Higuchi bound arises, let us use the planar coordinates
\be
\mathrm{d}s^2
=
a^2(\eta)
\left(
-\mathrm{d}\eta^2+\mathrm{d}\boldsymbol{x}^2
\right),
\qquad
a(\eta)=-\frac{1}{H\eta},
\qquad
\eta<0,
\ee
and decompose $h_{\mu\nu}$ into scalar, vector, and tensor components under spatial rotations \footnote{This standard scalar--vector--tensor decomposition of metric perturbations, together with the corresponding transversality and tracelessness conditions, is reviewed in detail in Refs.~\cite{Mukhanov2005,Maggiore2018,Weinberg2008}.
}
\begin{align}
h_{00}
&=
-2a^2\Phi, \quad
h_{0i}
=
a^2
\left(
\partial_iB+B_i^{\rm T}
\right),
\\
h_{ij}
&=
a^2
\left[
2\psi\,\delta_{ij}
+2\partial_i\partial_jE
+\partial_iE_j^{\rm T}
+\partial_jE_i^{\rm T}
+\gamma_{ij}^{\rm TT}
\right].
\end{align}
Here
\be
\partial_iB_i^{\rm T}=0,
\qquad
\partial_iE_i^{\rm T}=0,
\qquad
\partial_i\gamma_{ij}^{\rm TT}=0,
\qquad
\gamma_{ii}^{\rm TT}=0.
\ee
This is purely a kinematic decomposition and does not constitute a gauge choice. The transverse-traceless tensor $\gamma_{ij}^{\rm TT}$ carries the helicities $\pm2$, the transverse vector sector carries the helicities $\pm1$, and the scalar variables contain the helicity-zero mode. In the generic massive theory there is no gauge redundancy; instead, the non-dynamical variables impose the constraints required to leave five propagating degrees of freedom.

The crucial sector is the helicity-zero mode. After integrating out the non-dynamical scalar variables and reducing the action to a single propagating variable $v_0$, one obtains schematically~\cite{Higuchi1987}
\be
S_0^{(2)}
=
\frac12
\int d\eta\,d^3k\,
\mathcal{N}_0(\eta,k)
\left(
|v_0'|^2
-
\omega_0^2(\eta,k)|v_0|^2
\right),
\label{eq:helicity_zero_action}
\ee
where all other factors entering $\mathcal{N}_0$ are positive on the massive branch $m^2>0$, while
\be
\operatorname{sgn}\mathcal{N}_0
=
\operatorname{sgn}\left(m^2-2H^2\right).
\ee
The origin of the Higuchi bound is therefore transparent. For
\be
m^2>2H^2,
\ee
the helicity-zero mode has a positive kinetic term and positive norm. By contrast, in the interval
\be
0<m^2<2H^2,
\ee
its kinetic term has the opposite sign. The helicity-zero state is then a ghost, or equivalently a negative-norm state, and the corresponding de~Sitter representation is non-unitary~\cite{Higuchi1987}.

Something qualitatively different happens exactly at the boundary
\be
m^2=2H^2.
\ee
Here the vanishing of the helicity-zero kinetic coefficient should not be interpreted as an additional propagating zero-norm state. Instead, the theory develops the scalar gauge invariance
\be
\delta h_{\mu\nu}
=
\left(
\bar\nabla_\mu\bar\nabla_\nu
+H^2\bar g_{\mu\nu}
\right)\alpha,
\label{eq:PM-spin-two-gauge}
\ee
which removes the helicity-zero degree of freedom~\cite{DeserWaldron2001,DeserWaldron2001b}. The resulting \emph{partially massless} spin-two field therefore propagates four physical helicities,
\be
\lambda=\pm2,\ \pm1.
\ee
At the second special point, $m=0$, the scalar symmetry is enlarged to the usual linearized diffeomorphism invariance,
\be
\delta h_{\mu\nu}
=
\bar\nabla_\mu\xi_\nu+\bar\nabla_\nu\xi_\mu,
\ee
which removes both the helicity-zero and helicity-$\pm1$ sectors, leaving only the two graviton helicities $\lambda=\pm2$. The spin-two example thus provides a particularly concrete realization of the representation-theoretic discussion of \cref{subsec:dS-unitary-series}. For a compact summary of the unitary mass ranges and isolated unitary points for spin two, see \cref{fig:unitary-mass-spectrum}.

\subsection{Conformal weights and late-time behavior}
\label{subsec:conformal-weights}

A particularly useful consequence of the representation-theoretic
classification discussed above is that the same labels also determine how
fields behave near future conformal infinity, $\mathcal{I}^+$. Consider a
generic massive bosonic field of integer spin $s\geq1$, represented by a
totally symmetric rank-$s$ tensor $\sigma_{\nu_1\cdots\nu_s}$. For a generic massive representation, irreducibility requires the
transverse and traceless conditions
\be
\nabla^{\nu_1}\sigma_{\nu_1\cdots\nu_s}=0,
\qquad
\sigma^{\nu}{}_{\nu\nu_3\cdots\nu_s}=0.
\label{eq:spin_s_constraints}
\ee
These conditions project out the lower-spin components contained in a
generic symmetric tensor. The quadratic Casimir equation is then
equivalent to the covariant wave equation
\be
\left(
\Box-m_s^2
\right)
\sigma_{\nu_1\cdots\nu_s}=0,
\label{eq:spin_s_wave_equation}
\ee
where $\Box\equiv\nabla^\mu\nabla_\mu$ and
\be
m_s^2
\equiv
m^2-
\left(s^2-2s-2\right)H^2.
\label{eq:curvature-shifted-mass}
\ee
The difference between $m_s^2$ and the flat-space mass parameter $m^2$
is a curvature effect: for a spinning field, the de~Sitter Casimir acts
not only through the covariant Laplacian, but also on the tensor
indices~\cite{DeserWaldron2003,LeeBaumannPimentel2016}.

For our purposes, it is sufficient to focus on the maximal-helicity
components, $\lambda=\pm s$, which decouple from the lower-helicity
sectors. In the planar patch their mode functions obey
\be
\left[
\partial_\eta^2
+
\frac{2(s-1)}{\eta}\partial_\eta
+
k^2
+
\frac{1}{\eta^2}
\left(
\frac{m^2}{H^2}-2(s-1)
\right)
\right]
\sigma_s^{(\pm s)}(\eta,k)
=0,
\label{eq:spin-s-helicity-mode}
\ee
where $k=|\boldsymbol{k}|$~\cite{LeeBaumannPimentel2016}. It is
convenient to introduce
\be
\mu_s
\equiv
\sqrt{
\frac{m^2}{H^2}
-
\left(s-\frac12\right)^2
}.
\label{eq:mu-s-definition}
\ee
For the principal series $\mu_s$ is real, while for the complementary
series it is purely imaginary. The Bunch--Davies vacuum is selected by requiring that, sufficiently
far in the asymptotic past,
\be
-k\eta\gg1,
\qquad
\frac{k}{a}\gg H,m,
\ee
the physical wavelength is much shorter than all curvature and mass
scales, so that the canonically normalized mode approaches the
positive-frequency Minkowski solution~\cite{Bunch:1978yq}. For the maximal-helicity mode, the canonically
normalized variable is proportional to
\be
v_s^{(\pm s)}
=
a^{1-s}\sigma_s^{(\pm s)},
\ee
and the Bunch--Davies condition therefore reads
\be
v_s^{(\pm s)}(\eta,k)
\underset{-k\eta\to\infty}{\longrightarrow}
\frac{e^{-ik\eta}}{\sqrt{2k}}.
\label{eq:spin-s-BD-condition}
\ee

\paragraph{Conformal time and the Bunch--Davies vacuum.}
Let us pause here to take a closer look at the Bunch--Davies vacuum studied by T.~S.~Bunch and P.~C.~W.~Davies in
1978~\cite{Bunch:1978yq}, and the role of conformal time in its definition.
In conformal time, the spatially flat FLRW metric
$ds^2=a^2(\eta)(-d\eta^2+d\mathbf{x}^2)$ is manifestly conformal to
Minkowski space. In the asymptotic sub-horizon past, when mass and
curvature effects become negligible, the canonically normalized modes
recover Minkowski dynamics, and the Bunch--Davies prescription selects
the positive-frequency behavior
\cref{eq:spin-s-BD-condition}.
By contrast, in cosmic time,
$ds^2=-dt^2+a^2(t)d\mathbf{x}^2$, the scale factor multiplies only
the spatial part, so the conformal structure is not manifest:
the same positive-frequency phase is
$e^{-i\int^t [k/a(t')]\,dt'}$, rather than a fixed-frequency plane
wave in $t$. Thus, conformal time makes the vacuum prescription
particularly transparent, although the same state can equally
well be described using cosmic time.

\begin{figure}[t]
    \centering
    \includegraphics[width=0.6\linewidth]{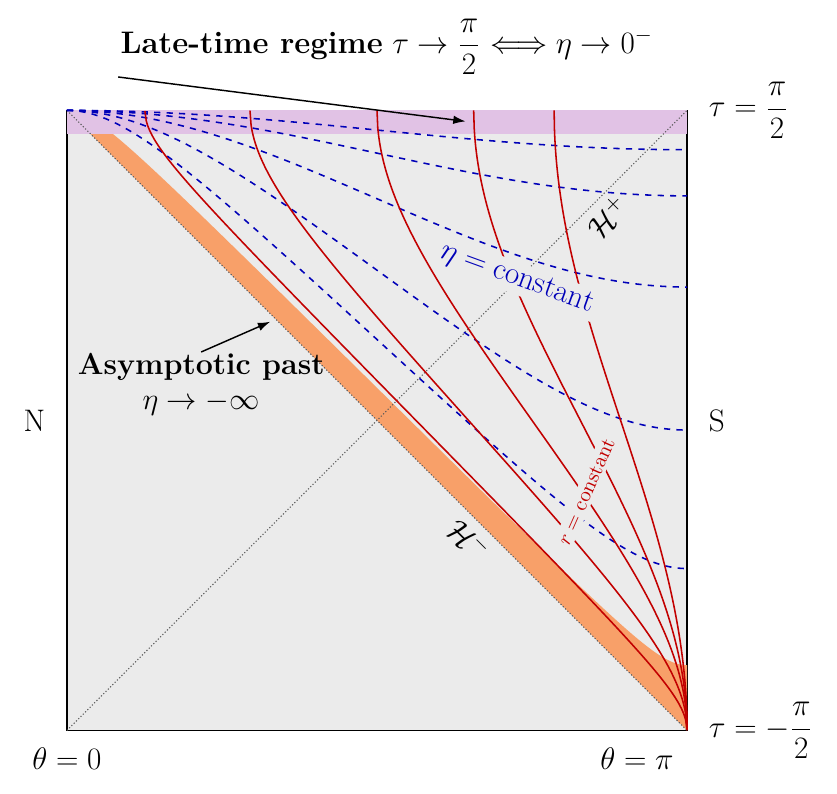}
   \caption{Same as \cref{fig:dS-penrose-3}, but with the late-time and asymptotic-past
regimes highlighted. The orchid-shaded band corresponds to the limit
$\eta\to0^{-}$, approaching future conformal infinity, while the orange-shaded
band adjacent to $\mathcal{H}^{-}$ marks the asymptotic past,
$\eta\to-\infty$.}
    \label{fig:Penrose-late-time}
\end{figure}

Using \cref{eq:eta-tau}, one sees that for any fixed finite comoving
momentum $k$, the Bunch--Davies limit $-k\eta\to\infty$ is obtained by
approaching the past null boundary of the expanding Poincar\'e patch,
\be
\sin\tau-\cos\theta\to0^{+}
\qquad\Longleftrightarrow\qquad
\tau\to\frac{\pi}{2}-\theta ,
\ee
from within the patch. Thus, in the Penrose diagram, the asymptotic
Minkowski regime in which the Bunch--Davies vacuum is defined lies
arbitrarily close to the null line $\mathcal H^{-}$, where
$k/a\gg H$ (and $k/a\gg m$ for a massive field). The late-time and asymptotic-past limits discussed above are illustrated in the
Penrose diagram presented in \cref{fig:Penrose-late-time}.
The corresponding solution of \cref{eq:spin-s-helicity-mode} is
\be
\sigma_s^{(\pm s)}(\eta,k)
=
\mathcal{N}_s \, 
(-k\eta)^{\frac32-s} \, 
H_{i\mu_s}^{(1)}(-k\eta),
\label{eq:spin-s-BD-mode}
\ee
where $H_{i\mu_s}^{(1)}$ is the Hankel function of the first kind.
The constant $\mathcal{N}_s$ fixes the overall canonical normalization
and phase of the mode. For unit-normalized maximal-helicity
polarization tensors, a convenient choice is
\be
\mathcal{N}_s
=
e^{i\pi/4}
e^{-\pi\mu_s/2}
\frac{\sqrt{\pi}}{2\sqrt{k}}
\left(\frac{k}{H}\right)^{s-1}.
\label{eq:spin-s-normalization}
\ee
Indeed, using the large-argument expansion
\be
\lim_{x\to\infty}H_{i\mu_s}^{(1)}(x)
=
\sqrt{\frac{2}{\pi x}}\,
e^{\pi\mu_s/2}
e^{i(x-\pi/4)},
\ee
one immediately recovers
\cref{eq:spin-s-BD-condition}. If a different normalization convention
is adopted for the polarization tensors, $\mathcal{N}_s$ is rescaled
accordingly, while the time dependence and all conclusions concerning
the conformal weights remain unchanged. See
Refs.~\cite{LeeBaumannPimentel2016,ArkaniHamedMaldacena2015} for
detailed discussions of massive spinning mode functions in de~Sitter
space.

The relation to the conformal weights becomes manifest in the late-time
limit. For a fixed comoving momentum, approaching $\mathcal{I}^+$
corresponds to $-k\eta\ll1$.
Using the small-argument behavior of the Hankel function,
we find
\be
\lim_{k\eta\to0^-}\sigma_s^{(\lambda)}(\eta,k)
=
\mathcal N_s
\left[
c_+(-k\eta)^{\frac32-s+i\mu_s}
+
c_-(-k\eta)^{\frac32-s-i\mu_s}
\right],
\qquad
\lambda=\pm s .
\ee
It is therefore natural to introduce the two conformal weights \cite{Strominger:2001pn}
\be
\Delta_\pm
=
\frac32\pm i\mu_s,
\label{eq:dS-conformal-weights}
\ee
which obey the mass--weight relation \cref{eq:mass-dimension-spin}. Therefore, the two roots satisfy
\be
\Delta_++\Delta_-=3,
\ee
and are related by the three-dimensional shadow transformation
$\Delta\leftrightarrow3-\Delta$~\cite{DeserWaldron2003,PethybridgeSchaub2022}.

More generally, the purely spatial coordinate components admit the
late-time expansion
\be
\lim_{\eta\to0^-} \sigma_{i_1\cdots i_s}(\eta,\boldsymbol{x})
=
(-H\eta)^{\Delta_+-s}
\sigma^+_{i_1\cdots i_s}(\boldsymbol{x})
+
(-H\eta)^{\Delta_--s}
\sigma^-_{i_1\cdots i_s}(\boldsymbol{x}).
\label{eq:late_time_spin_s}
\ee
The shift by $-s$ deserves some care. It is not a shift of the
conformal weight itself, but arises because
$\sigma_{i_1\cdots i_s}$ carries $s$ lower spatial coordinate indices.
Indeed, passing to a local orthonormal frame,
\be
e_{\hat\imath}{}^j
=
a^{-1}\delta_{\hat\imath}{}^j,
\ee
one obtains
\be
\sigma_{\hat\imath_1\cdots\hat\imath_s}
=
e_{\hat\imath_1}{}^{i_1}\cdots
e_{\hat\imath_s}{}^{i_s}
\sigma_{i_1\cdots i_s}
\sim
(-H\eta)^{\Delta_\pm}.
\label{eq:orthonormal-late-time}
\ee
Thus it is $\Delta_\pm$, rather than the coordinate-component
exponents $\Delta_\pm-s$, that characterize the intrinsic late-time
scaling of the field.

This also explains why the quantities $\Delta_\pm$ have the
interpretation of conformal dimensions. The planar-patch dilation
\be
\eta\to\lambda\eta,
\qquad
\boldsymbol{x}\to\lambda\boldsymbol{x},
\ee
is an exact de~Sitter isometry. At $\mathcal I^+$ it reduces to an
ordinary dilation of the three-dimensional boundary. The coefficients
$\sigma^\pm_{i_1\cdots i_s}(\boldsymbol{x})$ appearing in
\cref{eq:late_time_spin_s} therefore transform as three-dimensional
spin-$s$ conformal tensors of weights $\Delta_\pm$. In this sense, the
two independent bulk solutions furnish a pair of boundary
representations related by the shadow transformation
\cite{Strominger:2001pn,DeserWaldron2003,PethybridgeSchaub2022}.

The nature of the late-time behavior depends directly on the unitary
series discussed in the previous section. For the principal series,
$\mu_s\in\mathbb R$ in \cref{eq:dS-conformal-weights}. 
The two solutions have the same power-law envelope but opposite
logarithmic phases. In an orthonormal frame,
\be
\sigma_{\hat\imath_1\cdots\hat\imath_s}
\sim
(-H\eta)^{3/2}
e^{\pm i\mu_s\ln(-H\eta)}.
\label{eq:principal-late-time}
\ee
Using $-H\eta=e^{-Ht}$, this may equivalently be written as
\be
\sigma_{\hat\imath_1\cdots\hat\imath_s}
\sim
e^{-3Ht/2}
e^{\mp i\mu_sHt}.
\ee
Thus a principal-series field decays as $a^{-3/2}$ while oscillating in
cosmic time. In the heavy-mass limit, $m\gg H$, one has
$\mu_s\simeq m/H$, and the oscillation frequency approaches the
familiar flat-space value $m$. These logarithmic oscillations are also
responsible for the characteristic nonanalytic oscillatory signals of
massive particles in cosmological correlators
\cite{ArkaniHamedMaldacena2015,LeeBaumannPimentel2016}.

For the complementary series, it is convenient to write
\be
\mu_s=i\nu_s,
\qquad
\nu_s
=
\sqrt{
\left(s-\frac12\right)^2
-
\frac{m^2}{H^2}
},
\ee
so that the conformal weights become real,
\be
\Delta
=
\frac32\pm\nu_s.
\ee
The late-time behavior is then a sum of two ordinary power laws rather
than an oscillation,
\be
\sigma_{\hat\imath_1\cdots\hat\imath_s}
\sim
(-H\eta)^{\frac32-\nu_s}
\qquad\text{or}\qquad
(-H\eta)^{\frac32+\nu_s}.
\ee
The branch with the smaller conformal weight decays more slowly and
therefore dominates sufficiently close to $\mathcal I^+$.

\begin{figure}[t]
    \centering
    \includegraphics[width=0.6\linewidth]{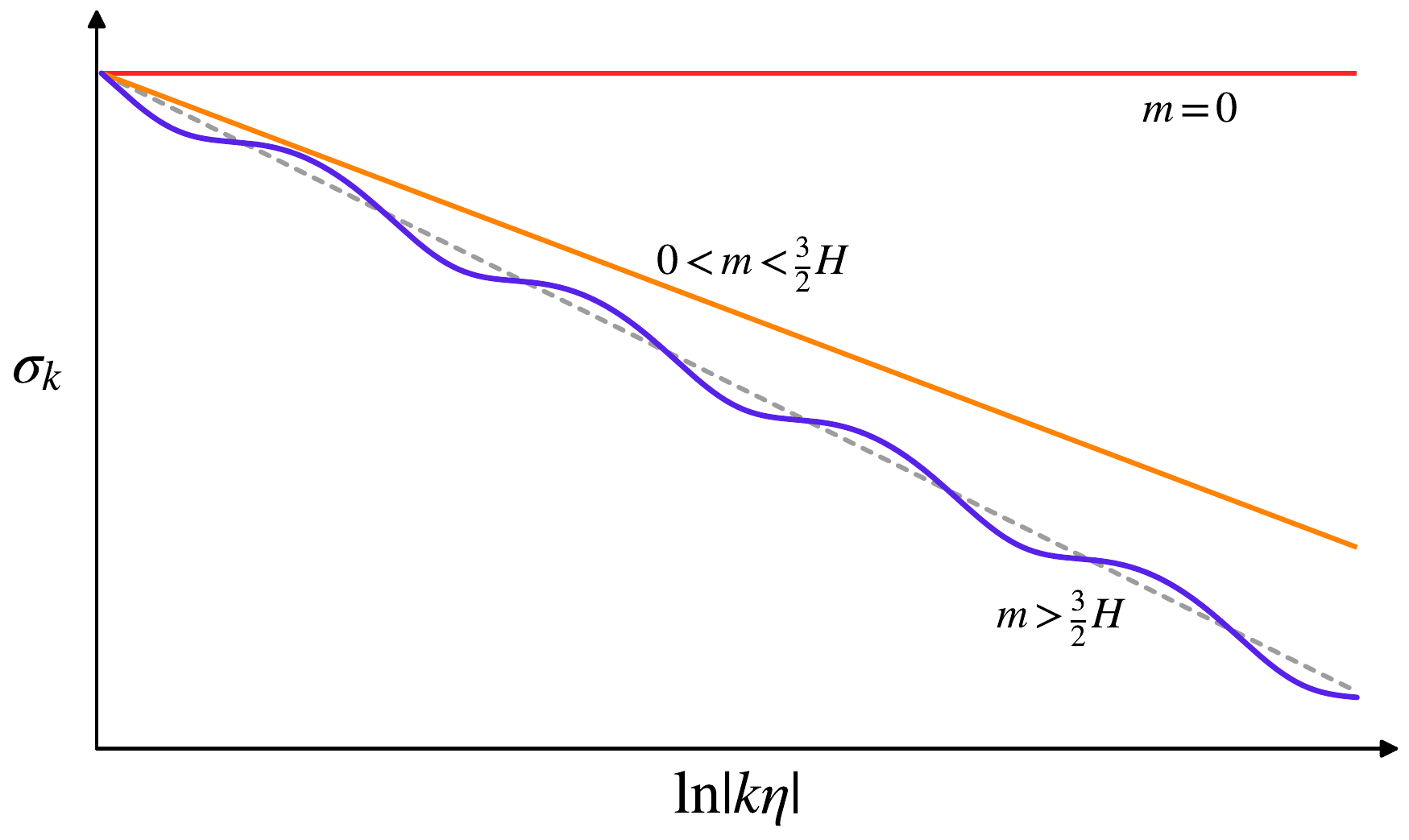}
    \caption{Schematic late-time evolution of a scalar Fourier mode
    $\sigma_k$ in de~Sitter spacetime. The massless mode, $m=0$, freezes
    to a constant, while a light field in the complementary series,
    $0<m<3H/2$, decays monotonically. For a heavy field in the principal
    series, $m>3H/2$, the mode exhibits logarithmic oscillations within
    a decaying power-law envelope, shown by the gray dotted curve. Adapted from  \cite{Baumann:2018muz} and modified for the present discussion.}
    \label{fig:massive-decay}
\end{figure}

At the isolated partially massless and strictly massless points, the
weights become integer-valued. With the depth convention
$t=0,\ldots,s-1$, introduced above, the corresponding masses satisfy
\be
\frac{m^2}{H^2}
=
s(s-1)-t(t+1),
\ee
and therefore
\be
\mu_s^2
=
-\left(t+\frac12\right)^2.
\ee
The two formal roots are
\be
\left\{
\Delta_+,\Delta_-
\right\}
=
\left\{
1-t,\,
2+t
\right\}.
\label{eq:PM-conformal-weights}
\ee
At these special points, however, the generic massive counting must be
supplemented by the additional gauge symmetry, which removes some of
the helicity components. The weight
\be
\Delta=2+t
\ee
is naturally associated with the corresponding partially conserved
spin-$s$ boundary operator, while $\Delta=1-t$ is its shadow
\cite{DeserWaldron2003,PethybridgeSchaub2022}. In particular, for the
strictly massless point $t=s-1$,
\be
\left\{\Delta_+,\Delta_-\right\}
=
\left\{
2-s,\,
s+1
\right\},
\ee
and the $\Delta=s+1$ branch has precisely the conformal dimension of a
conserved spin-$s$ current in three dimensions. The distinct late-time behavior of massless, complementary-series, and
principal-series scalar modes is illustrated schematically in
\cref{fig:massive-decay}.

Finally, the geometric meaning of this limit is especially transparent
in the Penrose diagram as we show in \cref{fig:Penrose-late-time}. Using \cref{eq:eta-tau}, the planar late-time
limit
\be
\eta\to0^- \quad \textmd{
maps to}
\quad
\tau\to\frac{\pi}{2},
\ee
namely to future conformal infinity $\mathcal I^+$ of the expanding
Poincar\'e patch. Hence the conformal weights $\Delta_\pm$ literally
measure how the physical field approaches the upper spacelike boundary
of the de~Sitter Penrose diagram. The central lesson is that the mass of a field in de~Sitter space is
encoded not only in its local oscillation frequency, as in Minkowski
space, but also in its scaling toward $\mathcal I^+$. The same
representation-theoretic data that determine the mass, spin, and
unitarity properties of the bulk field therefore also determine its
late-time conformal weights and, through them, its boundary behavior. \footnote{
The K\"all\'en--Lehmann representation in Minkowski and de~Sitter space follows from decomposing states into unitary irreducible representations of the isometry group
\cite{Kallen:1952zz,Lehmann:1954xi,Hogervorst2023Bootstrap,Loparco:2023KL}.
For invariant scalar two-point functions, the spectral parameter labels the quadratic Casimir, while positivity follows from the Hilbert-space norm. In de~Sitter space, analytic continuation of the spectral density can further encode late-time boundary dimensions
\cite{Hogervorst2023Bootstrap,DiPietro:2021sjt}; see also \cite{SalehiVaziri2024Boundary}.
}
 This imprint of bulk physics on late-time boundary data also underlies the
modern cosmological-correlator program.\footnote{The modern
cosmological-correlator, or cosmological-bootstrap, program treats late-time
correlators as fundamental observables, constraining them through de~Sitter
symmetry, locality, unitarity, and analyticity. Their non-analytic structure
can encode the masses and spins of particles present during inflation
\cite{Arkani-Hamed:2015bza,Baumann:2022jpr}.}

\section{Particle Production in de~Sitter Spacetime}
\label{sec:particle-production-dS}

We begin by introducing the Bogoliubov technique and the notion of particle production in the presence of time-dependent backgrounds that break time-translation symmetry. Quantum field theory in an expanding spacetime differs fundamentally from its flat-space counterpart because, in the absence of a preferred time-translation symmetry, there is in general no unique notion of positive frequency and hence no unique vacuum state. The Bogoliubov formalism provides a natural framework for comparing different choices of mode basis and for quantifying the mixing between positive- and negative-frequency components. This mixing is the basic mechanism underlying particle production in time-dependent backgrounds. In de~Sitter space, the cosmological expansion provides a particularly important realization of this phenomenon, with close connections to horizon thermality and to the quantum origin of cosmological perturbations. We first develop these ideas in a general setting and then apply them to tensor perturbations, before extending the discussion to scalar fields.

\subsection{Bogoliubov transformations and particle creation}
\label{subsec:bogoliubov-particle-creation}

The transformation now bearing Bogoliubov's name originated not in
cosmology, but in his 1947 microscopic theory of superfluidity. For a
weakly interacting Bose gas, he introduced linear combinations of
creation and annihilation operators to diagonalize the quadratic
Hamiltonian and identify its elementary quasiparticle
excitations~\cite{Bogoliubov1947}. The same canonical construction
provides a method for computing particle production in a time-dependent
background: the operators defining particles at late times need not
annihilate the state that was empty at early times.


\begin{figure}[t]
    \centering
    \includegraphics[width=0.95\textwidth]{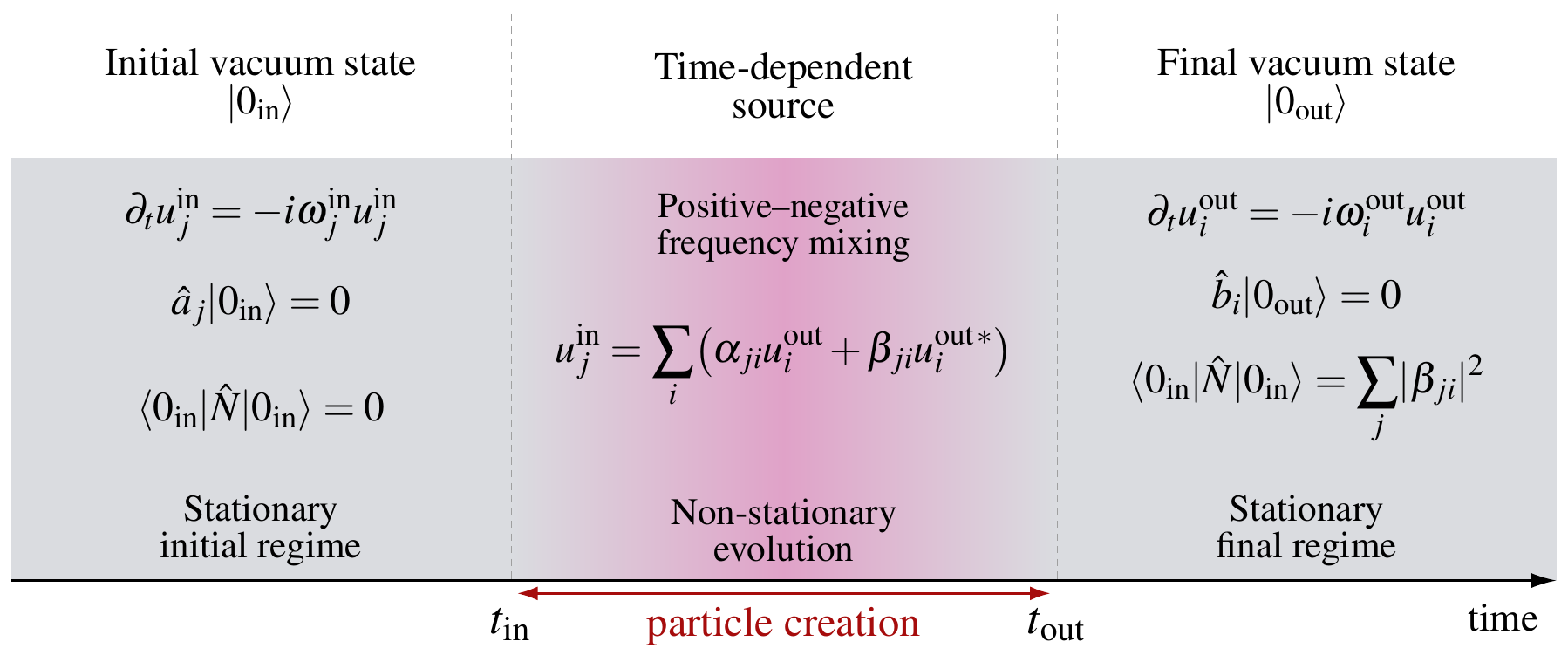}
    \caption{
    Bogoliubov transformation and particle production in a time-dependent background.
    In the asymptotic past and future, where the system becomes stationary, one can define
    positive-frequency mode bases and the corresponding in- and out-vacua.
    During the intermediate nonstationary evolution, positive- and negative-frequency modes
    mix according to
    $u_j^{\rm in}=\sum_i\!\left(\alpha_{ij}u_i^{\rm out}
    +\beta_{ij}u_i^{{\rm out}*}\right)$.
    A nonvanishing Bogoliubov coefficient $\beta_{ij}$ therefore implies that the initial
    vacuum contains particles with respect to the final particle basis,
    $\langle0_{\rm in}|\hat N_i^{\rm out}|0_{\rm in}\rangle
    =\sum_j|\beta_{ij}|^2$.
 Figure adapted and modified from \cite{Maleknejad:2023nyh}.
    }
    \label{fig:bogoliubov-particle-production}
\end{figure}

\paragraph{In- and out-particles.}
Consider a real scalar field $\varphi$ propagating in a prescribed
classical background $\mathcal S(t,\mathbf x)$. We work at quadratic
order in the field, so that its evolution is linear, and assume that
the background becomes stationary in the distant past and future.
The intervening time dependence may arise, for example, from a varying
effective mass or from the spacetime geometry itself. Here a
\emph{source} means a background entering the quadratic fluctuation
operator, rather than a term linear in $\varphi$.
The two stationary regimes supply positive-frequency mode bases,
$\{u_j^{\rm in}\}$ and $\{u_i^{\rm out}\}$, and corresponding particle
operators~\cite{Parker1969}:
\begin{equation}
\hat\varphi(x)
=\sum_j\left[\hat a_j u_j^{\rm in}(x)
             +\hat a_j^\dagger u_j^{{\rm in}*}(x)\right]
=\sum_i\left[\hat b_i u_i^{\rm out}(x)
             +\hat b_i^\dagger u_i^{{\rm out}*}(x)\right].
\label{eq:bog-field-expansions}
\end{equation}
The labels include all mode quantum numbers, and sums may be replaced
by integrals for continuous spectra. The associated reference vacua
are defined by (see \cref{fig:bogoliubov-particle-production})
\begin{equation}
\hat a_j|0_{\rm in}\rangle=0,
\qquad
\hat b_i|0_{\rm out}\rangle=0.
\label{eq:bog-vacua}
\end{equation}

Because both mode bases are complete, an initially positive-frequency
solution admits the late-time decomposition
\begin{equation}
u_j^{\rm in}
=\sum_i\left(\alpha_{ij}u_i^{\rm out}
             +\beta_{ij}u_i^{{\rm out}*}\right).
\label{eq:bog-mode-transformation}
\end{equation}
The coefficients $\beta_{ij}$ measure the admixture of negative
out-frequency. For the standard scalar kinetic term, the modes are
normalized with the conserved Klein--Gordon inner product \cite{BirrellDavies1982}
\begin{equation}
(u,v)_{\rm KG}
=i\int_\Sigma d\Sigma^\mu
\left[u^*\nabla_\mu v-(\nabla_\mu u^*)v\right],
\label{eq:bog-kg-product}
\end{equation}
where $d\Sigma^\mu$ is future directed. With
$(u_i,u_j)_{\rm KG}=\delta_{ij}$,
$(u_i^*,u_j^*)_{\rm KG}=-\delta_{ij}$, and
$(u_i,u_j^*)_{\rm KG}=0$, projection gives
\begin{equation}
\alpha_{ij}=(u_i^{\rm out},u_j^{\rm in})_{\rm KG},
\qquad
\beta_{ij}=-(u_i^{{\rm out}*},u_j^{\rm in})_{\rm KG}.
\label{eq:bog-projection}
\end{equation}
Thus the calculation consists of evolving positive-frequency initial
modes through the background and projecting them onto the final
positive- and negative-frequency bases.

Substituting \eqref{eq:bog-mode-transformation} into the field expansion
and collecting coefficients of $u_i^{\rm out}$ yields
\begin{equation}
\hat b_i
=\sum_j\left(\alpha_{ij}\hat a_j
             +\beta_{ij}^*\hat a_j^\dagger\right).
\label{eq:bog-operator-transformation}
\end{equation}
Consequently, the initial vacuum contains
\begin{equation}
\langle0_{\rm in}|\hat N_i^{\rm out}|0_{\rm in}\rangle
=\langle0_{\rm in}|\hat b_i^\dagger\hat b_i|0_{\rm in}\rangle
=\sum_j|\beta_{ij}|^2,
\label{eq:bog-particle-number}
\end{equation}
particles in the final mode $i$. In the Heisenberg picture, the state
remains $|0_{\rm in}\rangle$; what changes is the relation between the
initial and final particle operators. The out-vacuum is a reference
state, not the state into which the in-vacuum is assumed to evolve.
For a homogeneous, isotropic scalar background, the transformation
reduces to independent momentum pairs,
\begin{equation}
\hat b_{\mathbf k}
=\alpha_k\hat a_{\mathbf k}
 +\beta_k^*\hat a_{-\mathbf k}^\dagger,
\qquad
|\alpha_k|^2-|\beta_k|^2=1,
\qquad
\langle\hat N_{\mathbf k}^{\rm out}\rangle_{\rm in}=|\beta_k|^2.
\label{eq:bog-homogeneous}
\end{equation}
The opposite momenta reflect spatial translation invariance, while
the normalization follows from the bosonic commutation
relations.  \cref{fig:bogoliubov-particle-production} summarizes this in--out construction.

For more pedagogical discussions of the Bogoliubov formalism and particle
production in curved spacetime, see
Refs.~\cite{BirrellDavies1982,ParkerToms2009,FabbriNavarroSalas2005,Jacobson:2003vx}.
A celebrated application is Hawking's discovery that the formation of a black-hole event horizon leads to thermal particle creation, observed at future infinity as Hawking radiation \cite{Hawking:1975vcx}.
Particle production may also be induced by external gauge backgrounds, as in
the Schwinger effect \cite{Schwinger:1951nm}; related chiral phenomena in electromagnetic fields are
discussed in Ref.~\cite{Maleknejad:2023nyh}.

\paragraph{Vacuum choice without time-translation symmetry.}
Why does stationarity matter? A future-directed timelike Killing
vector $K^\mu$ that preserves the full background supplies a
geometrically preferred time evolution. Subject to stability and
appropriate boundary conditions, positive-frequency modes can be
selected by
\begin{equation}
i\mathcal L_K u_\omega=\omega u_\omega,
\qquad \omega>0,
\label{eq:bog-killing-frequency}
\end{equation}
and their annihilation operators define a natural ground-state vacuum.
A generic evolving spacetime has no such symmetry, so the geometry
alone need not select a unique division into positive and negative
frequencies. Merely choosing a time coordinate does not restore the
missing symmetry~\cite{HollandsWald2015}.

This is an absence of a \emph{preferred} vacuum, not an inability to
define quantum states. One can specify initial mode data and evolve
the resulting state. Physical admissibility is usually imposed through
the Hadamard condition, which fixes the universal short-distance
singularity structure of the two-point function but does not select a
unique state. Once physically motivated in- and out-bases are fixed,
\cref{eq:bog-particle-number} is a definite prediction; a unique
instantaneous particle number during the nonstationary interval need
not exist. In particular, introducing time-dependent or non-inertial coordinates in
Minkowski spacetime does not, by itself, imply particle
creation~\cite{HollandsWald2015,Jacobson:2003vx}. The Unruh effect provides
a classic illustration: a uniformly accelerated observer perceives the
Minkowski vacuum as a thermal state, with temperature
$T_{\rm U}=a/(2\pi)$, even though no particles are dynamically produced in
the inertial Minkowski description~\cite{Fulling:1972md,Unruh:1976db}.
The effect instead reflects the observer dependence of the notion of
positive frequency and hence of particles.

\subsection{Particle Creation by Cosmic Expansion}\label{sec:scalar-fields}

The connection between cosmic expansion and the mixing of positive- and
negative-frequency modes was already recognized by Schr\"odinger in
1939~\cite{Schrodinger1939}.  This was an important precursor,
but his wave-mechanical treatment was not a quantum-field calculation
of spontaneous particle production from the
vacuum~\cite{Schrodinger1939,Parker2012}.
In his 1966 doctoral work and subsequent papers of 1968–69, Parker quantized a field in a universe with stationary asymptotic regions, showing that cosmic expansion evolves the in-vacuum into a particle-containing out-state. These limits define particles unambiguously despite the absence of time-translation symmetry during the expansion~\cite{Parker1966,Parker1968,Parker1969}. Here we present only the essence of this mechanism; for detailed treatments, see \cite{ParkerToms2009,Kolb:2023ydq}.

\begin{figure}[t]
    \centering
    \includegraphics[width=\textwidth]
    {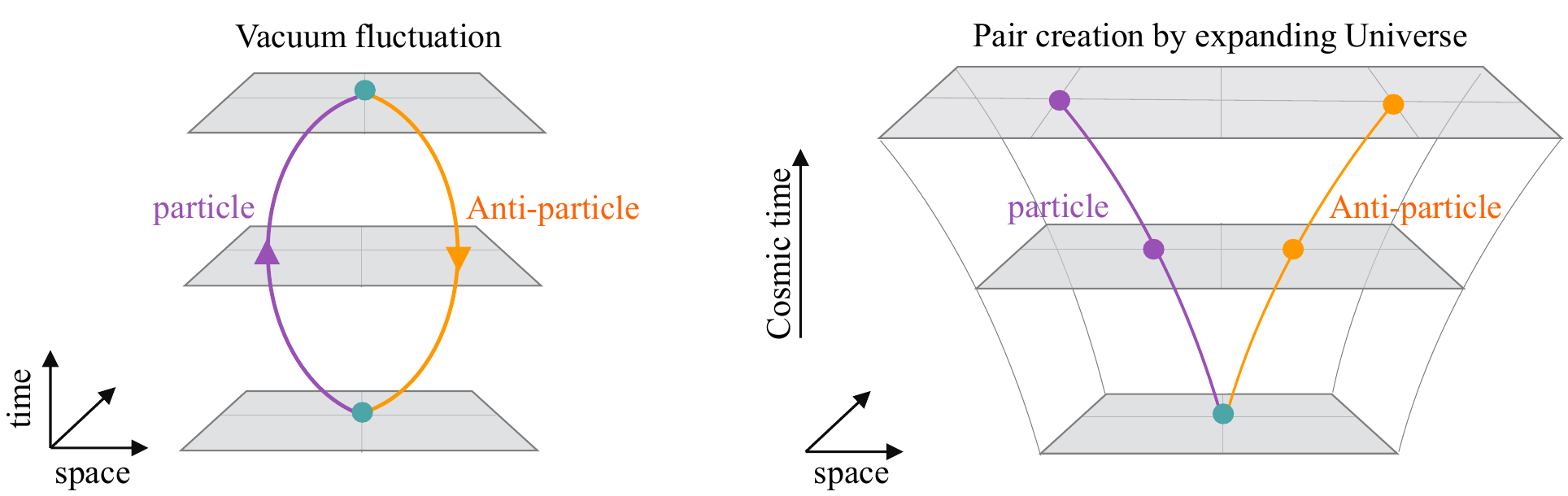}
    \caption{Vacuum fluctuations versus cosmological particle production.
    \textbf{Left:} A virtual particle--antiparticle pair fluctuates out of and back
    into the vacuum in a stationary spacetime, represented by spatial
    hypersurfaces of fixed size.
    \textbf{Right:} In an expanding Universe, the time dependence of the background
    can convert vacuum fluctuations into real particles whose worldlines
    persist across successive expanding spatial hypersurfaces.
    The coloured dots mark the particle and antiparticle at different
    times. The illustration is schematic.}
    \label{fig:vacuum-fluctuation-particle-production}
\end{figure}

Consider the spatially flat FLRW
metric in four dimensions,
\begin{equation}
ds^2=-dt^2+a^2(t)d\mathbf x^2
=a^2(\eta)\left(-d\eta^2+d\mathbf x^2\right),
\qquad d\eta=\frac{dt}{a(t)},
\label{eq:bog-flrw}
\end{equation}
and a free scalar with action
\begin{equation}
S_\varphi=-\frac12\int d^4x\,\sqrt{-g}
\left[g^{\mu\nu}\partial_\mu\varphi\partial_\nu\varphi
      +(m^2+\xi R)\varphi^2\right].
\label{eq:bog-scalar-action}
\end{equation}
Here $\xi$ is the curvature coupling, with $\xi=0$ for minimal
coupling and $\xi=1/6$ for conformal coupling. Writing the canonically
rescaled mode as $v_k=a\varphi_k$, and using
$R=6a''/a^3$, gives
\begin{equation}
v_k''+\Omega_k^2(\eta)v_k=0,
\qquad
\Omega_k^2(\eta)
=\underbrace{k^2+a^2m^2+(6\xi-1)\frac{a''}{a}}_{\textmd{time-varing}}.
\label{eq:bog-cosmological-oscillator}
\end{equation}
Primes denote conformal-time derivatives, and canonical normalization
requires $v_kv_k^{*\prime}-v_k'v_k^*=i$. Each Fourier mode is therefore
a quantum oscillator whose frequency is controlled by the evolving
geometry~\cite{Parker1969,KolbLong2024}.

For the cosmological (expanding Poincaré) patch of de Sitter,
\begin{equation}
a(\eta)=-\frac{1}{H\eta},
\qquad
-\infty<\eta<0,
\end{equation}
and the mode equation becomes
\begin{equation}
v_k''
+
\left[
k^2-\frac{\nu^2-\frac14}{\eta^2}
\right]v_k=0,
\qquad
\nu^2=\frac94-\frac{m^2}{H^2}-12\xi .
\end{equation}
As discussed in \cref{subsec:conformal-weights} and \cref{eq:spin-s-BD-condition}, the Bunch--Davies mode is selected by the positive-frequency condition in the asymptotic past,
\begin{equation}
v_k^{\rm BD}(\eta)
\longrightarrow
\frac{e^{-ik\eta}}{\sqrt{2k}},
\qquad
\eta\to-\infty .
\end{equation}
Near the future boundary, it instead behaves as
\begin{equation}
v_k^{\rm in}(\eta)
\longrightarrow
\frac{1}{\sqrt{2H\mu}}
\left[
\underbrace{
\alpha_k(-H\eta)^{\frac12+i\mu}
}_{\text{positive frequency}}
+
\underbrace{
\beta_k(-H\eta)^{\frac12-i\mu}
}_{\text{negative frequency}}
\right],
\qquad
\eta\to0^- .
\label{eq:bog-ds-out}
\end{equation}
with $\mu=-i\nu$, and
\begin{equation}
(-H\eta)^{\pm i\mu}
=e^{\mp i\mu Ht},
\end{equation}
so the two terms have positive and negative frequency with respect to cosmic time. Canonical normalization implies
\begin{equation}
|\alpha_k|^2-|\beta_k|^2=1,
\end{equation}
and the late-time occupation number is \(n_k=|\beta_k|^2\). Hence the physical number density,
\begin{equation}
n_{\rm out}
=\frac{1}{a_{\rm out}^3}
\int\frac{d^3k}{(2\pi)^3}\,|\beta_k|^2.
\label{eq:bog-number-density}
\end{equation}
No non-gravitational interaction is required: the time dependence of
the metric itself changes the quadratic field
Hamiltonian~\cite{Parker1968,KolbLong2024}.

Particle creation is not, however, an automatic consequence of any
change in $a$. In an oscillatory regime with $\Omega_k^2>0$, a slowly
varying frequency admits the adiabatic approximation
\begin{equation}
v_k^{\rm ad}(\eta)\simeq
\frac{1}{\sqrt{2\Omega_k(\eta)}}
\exp\!\left[-i\int^\eta
\Omega_k(\tilde\eta)\,d\tilde\eta\right],
\qquad
\left|\frac{\Omega_k'}{\Omega_k^2}\right|\ll1,
\label{eq:bog-adiabatic}
\end{equation}
together with the corresponding higher-derivative conditions. In a
smooth adiabatic limit, frequency mixing is suppressed; departures
from adiabatic evolution can generate nonzero
$\beta_k$~\cite{Parker2012}.
An exact exception is a massless, conformally coupled scalar:
\begin{equation}
m=0,\qquad \xi=\frac16
\quad\Longrightarrow\quad
\Omega_k^2=k^2,\qquad \beta_k=0,
\label{eq:bog-conformal-exception}
\end{equation}
for the conformal-vacuum choice. The same conclusion holds for the free Maxwell field in four dimensions. More generally, because spatially flat FLRW spacetime is conformally related to Minkowski space, free conformally invariant fields prepared in the conformal vacuum are not produced by the homogeneous expansion at tree level \cite{ParkerToms2009}.\footnote{Cosmological perturbations can evade this protection. In particular, tensor perturbations render the geometry non-conformally flat, allowing a stochastic gravitational-wave background to produce massless fermions at gravitational one loop~\cite{Maleknejad:2024ybn,Maleknejad:2024hoz,Garani:2025qnm}. If the background is chiral, it can also generate a lepton asymmetry through the gravitational chiral anomaly \cite{Alexander:2004us,Maleknejad:2014wsa,Maleknejad:2016dci,Caldwell:2017chz,Alexander:2018fjp,Maleknejad:2024vvf}.}

For sufficiently heavy fields, the Bogoliubov coefficient takes the approximate form
\begin{equation}
|\beta_k|^2 \simeq e^{-2\pi m/H},
\end{equation}
revealing an exponential suppression characteristic of non-perturbative particle production. This spectrum admits a thermal interpretation, corresponding to a temperature \cite{GibbonsHawking1977}
\begin{equation}
T_{\rm dS} = \frac{H}{2\pi}.
\end{equation}
For light fields, however, the notion of particles becomes ambiguous, when the late-time evolution is non-oscillatory. In this regime, the physically relevant description is in terms of squeezed quantum states and the emergence of classical stochastic fluctuations rather than particle quanta.

\subsection{Cosmological Perturbations in de~Sitter}
\label{subsec:tensor-scalar-perturbations}

Cosmological perturbations in an expanding Universe constitute a vast and long-established subject that warrants a dedicated treatment. Here we only scratch the surface, restricting our attention to linear tensor and scalar perturbations around exact de~Sitter spacetime in \(3+1\) dimensions. Consider the expanding Poincar\'e patch of de~Sitter,
\begin{equation}
ds^2
=
a^2(\eta)
\left[
-d\eta^2
+
\bigl(\delta_{ij}+h_{ij}\bigr)dx^i dx^j
\right],
\qquad
a(\eta)=-\frac{1}{H\eta},
\label{eq:ds-perturbed-metric}
\end{equation}
in which $h_{ij}$ denotes the tensor perturbation of the metric, corresponding
to gravitational waves. The tensor perturbation is transverse and traceless,
\begin{equation}
\partial_i h_{ij}=0,
\qquad
h_{ii}=0,
\label{eq:tensor-tt-conditions}
\end{equation}
which can be decomposed into its two linear polarization states
\begin{equation}
h_{ij}(\eta,\mathbf{x})
=
\sum_{\lambda=+,\times}
\int\frac{d^3k}{(2\pi)^3}
\left[
e_{ij}^{\lambda}(\hat{\mathbf{k}})
h_k^\lambda(\eta)
e^{i\mathbf{k}\cdot\mathbf{x}}
+\mathrm{c.c.}
\right].
\label{eq:tensor-polarization-decomposition}
\end{equation}
Here $e_{ij}^{+}$ and $e_{ij}^{\times}$ denote the plus and cross
polarization tensors, respectively. 
Expanding the Einstein--Hilbert action,
\begin{equation}
S_{\rm EH}
=
\frac{M_{\rm Pl}^2}{2}
\int d^4x\,\sqrt{-g}\,R,
\end{equation}
to quadratic order in the transverse-traceless perturbation $h_{ij}$,
and discarding boundary terms, gives
\begin{equation}
S_h^{(2)}
=
\frac{M_{\rm Pl}^2}{8}
\int d\eta\,d^3x\,a^2
\left[
h_{ij}'h_{ij}'
-
\partial_\ell h_{ij}\partial_\ell h_{ij}
\right].
\label{eq:tensor-quadratic-action}
\end{equation}
Choosing the polarization normalization
\begin{equation}
e_{ij}^{\lambda}(\hat{\mathbf{k}})
e_{ij}^{\lambda' *}(\hat{\mathbf{k}})
=
\delta^{\lambda\lambda'},
\end{equation}
and introducing the canonically normalized mode
\begin{equation}
v_k^\lambda
=
\frac{aM_{\rm Pl}}{2}\,h_k^\lambda,
\label{eq:tensor-canonical-variable}
\end{equation}
the quadratic action becomes
\begin{equation}
S_h^{(2)}
=
\frac{1}{2}
\sum_{\lambda=+,\times}
\int d\eta\,\frac{d^3k}{(2\pi)^3}
\left[
\left|(v_k^\lambda)'\right|^2
-
\left(
k^2-\frac{a''}{a}
\right)
\left|v_k^\lambda\right|^2
\right].
\label{eq:tensor-canonical-action}
\end{equation}
The corresponding mode equation is therefore
\begin{equation}
(v_k^\lambda)''
+
\left(
k^2-\frac{a''}{a}
\right)v_k^\lambda
=
0,
\label{eq:tensor-mode-equation}
\end{equation}
where $ \frac{a''}{a}=\frac{2}{\eta^2}$.  Imposing the Bunch--Davies condition \cref{eq:spin-s-BD-condition} in the asymptotic past gives
\begin{equation}
v_k^\lambda(\eta)
=
\frac{e^{-ik\eta}}{\sqrt{2k}}
\left(
1-\frac{i}{k\eta}
\right).
\label{eq:tensor-bd-mode}
\end{equation}
The mode initially oscillates as a Minkowski vacuum fluctuation for
$-k\eta\gg1$, but freezes after horizon exit
\begin{equation}
h_k^\lambda(\eta)
\longrightarrow
\frac{\sqrt{2}\,H}{M_{\rm Pl}k^{3/2}},
\qquad
-k\eta\to0.
\label{eq:tensor-superhorizon-limit}
\end{equation}
Summing over the two polarizations yields the scale-invariant tensor
power spectrum,
\begin{equation}
\mathcal{P}_h(k)
\equiv
\sum_{\lambda}
\frac{k^3}{2\pi^2}
\left|h_k^\lambda\right|^2
=
\frac{2H^2}{\pi^2M_{\rm Pl}^2}.
\label{eq:tensor-power-spectrum}
\end{equation}

The evolution of scalar fields in de~Sitter spacetime was discussed in \cref{sec:scalar-fields}.
 Here we specialize to minimal coupling,
$\xi=0$, and to the effectively light inflaton fluctuation
$\delta\phi$. In the super-horizon limit, the Bunch--Davies mode freezes,
up to an irrelevant phase:
\begin{equation}
\delta\phi_k(\eta)
\longrightarrow
\frac{H}{\sqrt{2k^3}},
\qquad
-k\eta\ll1.
\label{eq:scalar-superhorizon-limit}
\end{equation}
Consequently,
\begin{equation}
\mathcal{P}_{\delta\phi}(k)
=
\frac{k^3}{2\pi^2}
\left|\delta\phi_k\right|^2
=
\left(
\frac{H}{2\pi}
\right)^2.
\label{eq:inflaton-fluctuation-spectrum}
\end{equation}

During slow-roll inflation, this field fluctuation is converted into the
curvature perturbation,
\begin{equation}
\zeta
=
-\frac{H}{\dot{\bar{\phi}}}\,\delta\phi,
\qquad
\mathcal{P}_\zeta(k)
=
\frac{H^2}
{8\pi^2\epsilon M_{\rm Pl}^2},
\qquad
\epsilon
=
-\frac{\dot H}{H^2}.
\label{eq:curvature-perturbation-spectrum}
\end{equation}
In single-field inflation, $\zeta$ is conserved on super-horizon scales. During cosmic inflation, quantum fluctuations in the scalar and tensor sectors are stretched to super-horizon scales, leaving potentially observable imprints in the CMB and large-scale structure. The measured scalar spectrum is in remarkable agreement with inflationary predictions, while primordial tensor modes remain undetected.\footnote{Planck finds a nearly scale-invariant scalar spectrum with $n_s=0.9649\pm0.0042$ at $68\%$ confidence \cite{Planck:2018jri}. The BICEP/Keck BK18 analysis finds no evidence for primordial tensor modes and constrains the tensor-to-scalar ratio to $r_{0.05}<0.036$ at $95\%$ confidence \cite{BICEP:2021xfz}.} 
For comprehensive treatments of cosmological perturbations during inflation, see \cite{Weinberg2008,Baumann:2018muz}; for a comprehensive treatment of the mathematical formulation, cosmological evolution, and observational prospects of gravitational waves, see
\cite{Maleknejad:2025clz}.

In exact
de~Sitter, by contrast, $\dot{\bar{\phi}}=0$, so this conversion is
ill-defined.  The essential distinction may be summarized as
\begin{equation}
\begin{aligned}
\underbrace{\dot{\bar{\phi}}\neq0}_{\text{inflationary clock}}
&\quad\Longrightarrow\quad
\zeta=-\frac{H}{\dot{\bar{\phi}}}\,\delta\phi,
\qquad
\dot{\zeta}\simeq0
\quad (k\ll aH),
\\[1ex]
\underbrace{\dot{\bar{\phi}}=0}_{\text{exact de~Sitter}}
&\quad\Longrightarrow\quad
\delta\phi\nrightarrow\zeta
\qquad
\text{(no dynamical clock)}.
\end{aligned}
\label{eq:inflationary-clock}
\end{equation}
In single-field inflation, a rolling background converts field fluctuations
into a conserved super-horizon curvature perturbation, whereas exact de~Sitter provides no clock for this conversion. More generally, for
sufficiently light scalar fields, the slow decay of super-horizon modes may
render correlation functions sensitive to long-wavelength
fluctuations.\footnote{Infrared effects in de~Sitter space are field- and
observable-dependent. Infrared-enhanced or secular terms do not, by themselves,
imply an instability: depending on the setting, they may encode long-wavelength
dynamics requiring re-summation, or cancel from suitably defined observables
\cite{Anninos:2014lwa,Gorbenko:2019rza,Cohen:2020php,Senatore:2012nq}.}

\subsection{Cosmic No-Hair and the Fate of Anisotropies in de~Sitter}
\label{sec:cosmic-no-hair}

\begin{figure}[t]
    \centering
    \includegraphics[width=\textwidth]{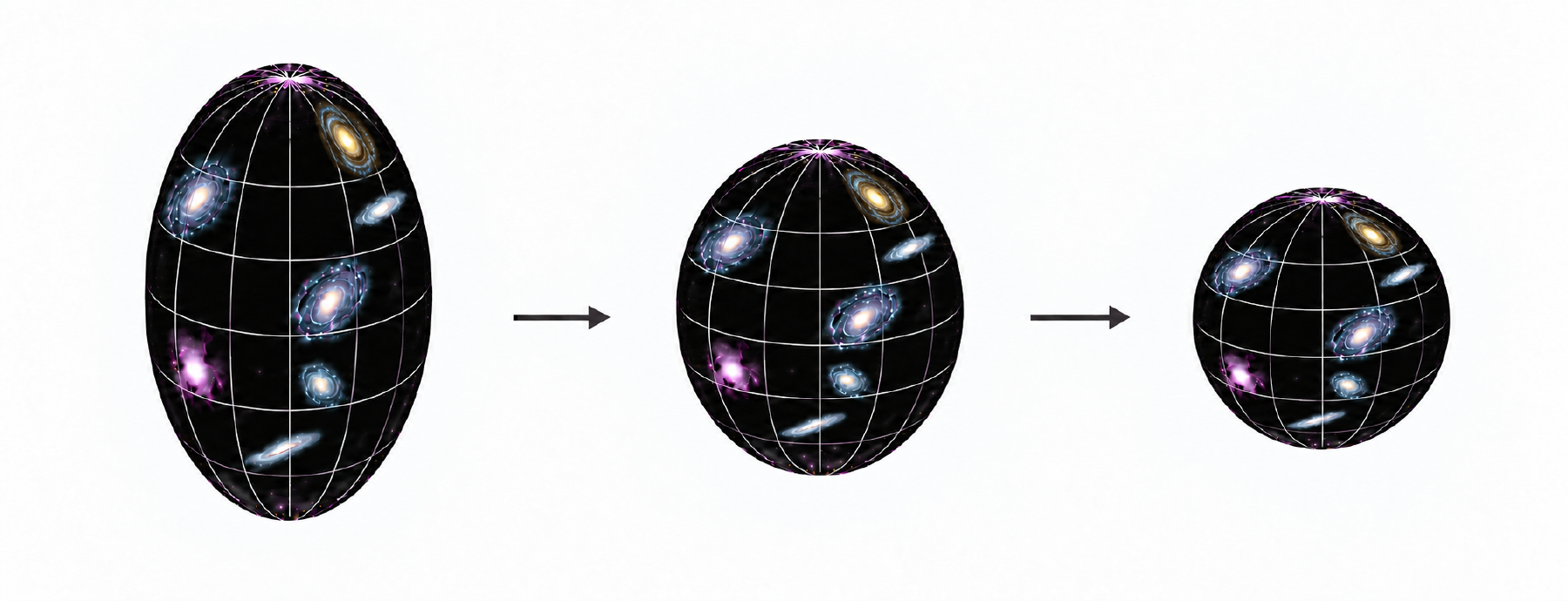}
    \caption{Cosmic no-hair and the fate of anisotropies. A schematic illustration of the evolution from an anisotropic geometry toward
an isotropic de~Sitter state. Observationally, CMB measurements are consistent
with statistical isotropy and place stringent bounds on homogeneous anisotropic
expansion \cite{Planck2018Isotropy}. The central question addressed
by the cosmic no-hair conjecture is whether this observed near-isotropy admits
a dynamical explanation: can accelerated expansion erase primordial
anisotropies and drive a generic expanding universe toward an isotropic
de~Sitter attractor?}
    \label{fig:cosmic-no-hair}
\end{figure}

The particle-production analysis above treats de~Sitter spacetime as a prescribed background. Since the mixing of positive- and negative-frequency modes is controlled by the evolving geometry, it is important to ask whether an expanding Universe is dynamically driven toward de~Sitter or whether initial anisotropy, spatial curvature and matter sources can persist and modify this evolution. The fate of such departures from de~Sitter is therefore a natural question with which to conclude our discussion. 
The \textit{cosmic no-hair conjecture} expresses the idea that accelerated expansion erases the memory of sufficiently generic initial conditions: anisotropy, spatial curvature and ordinary matter become dynamically negligible, and the local geometry approaches de Sitter space (see \cref{fig:cosmic-no-hair}). The terminology is inspired by the analogous ``no-hair'' property of black holes \cite{MisnerThorneWheeler1973}. The underlying intuition appeared already in the work of Gibbons and Hawking and of Hawking and Moss \cite{GibbonsHawking1977,HawkingMoss1982}, and was made precise for homogeneous cosmologies by Wald \cite{Wald1983}.

\paragraph{Cosmic no-hair theorem (1983):} Wald’s  cosmic no-hair theorem \cite{Wald1983} established that a broad class of expanding Bianchi cosmologies approach de Sitter spacetime.  To see the essential physics, consider a spatially homogeneous, but not necessarily isotropic, Bianchi geometry,
\begin{equation}
ds^2=-dt^2+h_{ij}(t)\,\omega^i\omega^j,
\end{equation}
where \(\omega^i\) are the invariant spatial one-forms associated with the Bianchi symmetry group, satisfying
\begin{equation}
d\omega^i=-\frac12 C^{i}_{~jk}\,\omega^j\wedge\omega^k ,
\end{equation}
with \(C^{i}_{~jk}\) the structure constants of the corresponding Bianchi type. All time dependence of the spatial metric is therefore contained in \(h_{ij}(t)\).  For completeness, a brief review of the Bianchi family of spatially
homogeneous cosmologies is provided in \cref{sec:bianchi}.

Let $n^\mu$ be the unit tangent vector field of the congruence of timelike geodesics orthogonal to the homogeneous space-like hypersurfaces $\Sigma_t$. Then, we obtain the following covariant form for the spatial metric $h_{\mu\nu}$,
\begin{equation}
h_{\mu\nu}=g_{\mu\nu}+n_\mu n_\nu ,
\end{equation}
where $g_{\mu\nu}$ is the metric of the space-time. The extrinsic curvature of $\Sigma_t$ is defined as
\begin{equation}
K_{\mu\nu}\equiv \frac{1}{2}\mathcal{L}_n h_{\mu\nu}
=\frac{1}{2}\dot{h}_{\mu\nu},
\end{equation}
where the dot represents derivative with respect to the proper time $t$. One can decompose $K_{\mu\nu}$ into trace and trace-free parts,
\begin{equation}
K_{\mu\nu}=\frac{1}{3}K h_{\mu\nu}+\sigma_{\mu\nu},
\end{equation}
where \(K=3H\) is the volume expansion and \(\sigma_{\mu\nu}\) is the shear.  The dimensionless quantity
\begin{equation}
\frac{\sigma}{H},
\qquad
(\sigma^2\equiv\sigma_{ij}\sigma^{ij}),
\end{equation}
therefore measures the importance of anisotropic expansion.

Now, consider Einstein gravity in units \(8\pi G=1\),  with
\begin{equation}
T_{\mu\nu}
=
-\Lambda g_{\mu\nu}
+\widetilde T_{\mu\nu},
\qquad \Lambda>0 ,
\end{equation}
where \(\Lambda\) is constant and \(\widetilde T_{\mu\nu}\) satisfies the dominant and strong energy conditions. Wald showed that every initially expanding Bianchi universe, except for the generic Bianchi IX case,\footnote{Bianchi IX can also isotropize when the cosmological constant is sufficiently large compared with the positive spatial-curvature contribution \cite{Wald1983}.} approaches de Sitter exponentially rapidly. The proof is remarkably economical. The Hamiltonian constraint gives, for all Bianchi types other than IX,
\begin{equation}
K^2-3\Lambda
=
3 T_{\mu\nu}n^\mu n^\nu
-\frac32\,{}^{(3)}R
+\frac32\,\sigma^2
\geq0 ,
\end{equation}
since $T_{\mu\nu}n^\mu n^\nu\geq 0$ and \({}^{(3)}R\leq0\). Hence
\begin{equation}
K\geq\sqrt{3\Lambda}.
\end{equation}
Meanwhile the Raychaudhuri equation, together with the strong energy condition, implies
\begin{equation}
\dot K
\leq
\Lambda-\frac{K^2}{3}.
\end{equation}
Thus \(K\) is driven towards the fixed point
\begin{equation}
K\longrightarrow \sqrt{3\Lambda},
\qquad
\underbrace{H(t)}_{\rm expanding\ geometry}
\longrightarrow
\underbrace{\sqrt{\frac{\Lambda}{3}}}_{\rm de\ Sitter}.
\end{equation}
on a time scale of order \(H_\Lambda^{-1}\). The constraint then forces
\begin{equation}
\frac{\sigma}{H}\longrightarrow0,
\qquad
\frac{{}^{(3)}R}{H^2}\longrightarrow0,
\qquad
\frac{\widetilde\rho}{H^2}\longrightarrow0 .
\end{equation}
In this precise sense, de Sitter space is the late-time attractor: the universe loses its homogeneous classical ``hair.''

\paragraph{Revisiting cosmic no-hair theorem for inflation (2012):} Maleknejad and Sheikh-Jabbari \cite{MaleknejadSheikhJabbari2012} pointed out that realistic inflationary matter does not satisfy the energy conditions underlying Wald’s theorem, and showed that inflation can therefore evade the standard cosmic no-hair conjecture. Such deviations from isotropy are nevertheless controlled by the departure of the inflationary background from exact de Sitter expansion, quantified by the slow-roll parameters. In the following, we briefly review their formulation and the resulting bounds on anisotropic hair. Inflation is only approximately de Sitter and must eventually end, i.e.
\begin{equation}
\underbrace{\epsilon\equiv-\frac{\dot H}{H^2}>0}_{\text{Inflation is quasi-de Sitter}}, \quad \text{while} \quad \underbrace{\epsilon_{\rm dS}=0}_{\text{exact de Sitter}}.
\end{equation}
Moreover, matter responsible for inflation need not satisfy the energy-condition assumptions imposed on \(\widetilde T_{\mu\nu}\), e.g. models containing  gauge fields and anisotropic stress  \cite{WatanabeKannoSoda2009,Maleknejad:2011jr,Maleknejad:2013npa,Adshead:2018emn,Wolfson:2020fqz,Maleknejad:2020yys,Wolfson:2021fya}.

A generalized cosmic no-hair theorem appropriate to realistic inflationary 
settings was formulated in \cite{MaleknejadSheikhJabbari2012}. The analysis 
revisits Wald's theorem by identifying the energy conditions relevant during 
inflation and determining the fate of anisotropic hair in generic Bianchi 
cosmologies. In this setting, the energy-momentum tensor can be decomposed as
\begin{equation}
T_{\mu\nu}
=
-\Lambda(t) g_{\mu\nu}
+\mathcal{T}_{\mu\nu},
\end{equation}
where the time-dependent vacuum-energy component, $\Lambda(t)$, drives the 
quasi-de Sitter expansion, while $\mathcal{T}_{\mu\nu}$ describes departures 
from an exact cosmological constant. With respect to the comoving timelike 
congruence $n^\mu$, the latter may be written as
\begin{equation}
\mathcal{T}_{\mu\nu}
=
\rho\, n_\mu n_\nu
+P\, h_{\mu\nu}
+\pi_{\mu\nu},
\qquad
\pi^\mu{}_\mu=0,
\qquad
\pi_{\mu\nu}n^\nu=0 ,
\end{equation}
where $\rho$ and $P$ denote the effective energy density and isotropic pressure,
and $\pi_{\mu\nu}$ is the anisotropic stress. The essential departure from Wald's setting is that inflationary matter need 
not satisfy the strong energy condition. Instead, the stress tensor is naturally 
separated into a dominant, slowly varying cosmological-constant-like component 
and a residual contribution $\mathcal{T}_{\mu\nu}$, on which the appropriate 
energy conditions are imposed. This formulation makes it possible to determine,
without assuming exact de Sitter expansion, whether anisotropic hair is erased,
survives, or is continuously sourced during inflation.

The shear is then sourced according to
\begin{equation}
\dot\sigma^i{}_j
+3H\sigma^i{}_j
+{}^{(3)}S^i{}_j
=
\pi^i{}_j .
\label{eq:shear-nohair}
\end{equation}
After a few e-folds the spatial-curvature term and the homogeneous solution of \eqref{eq:shear-nohair} are exponentially suppressed; any surviving anisotropy is therefore controlled by the anisotropic stress. The important result is that inflation itself places a model-independent bound on this hair:
\begin{equation}
\frac{\sigma_{\mu\nu}\sigma^{\mu\nu}}{2H^2}
\leq\epsilon(t),
\qquad
\frac{|\pi^i{}_j|}{H^2}
\leq4\epsilon(t) ,
\label{eq:inflationary-nohair-bound}
\end{equation}
and hence, componentwise,
\begin{equation}
\frac{|\sigma^i{}_j|}{H}
\lesssim\sqrt{2\epsilon(t)}, \quad \text{(slow-roll bound on anisotropic hair).}
\end{equation}
Thus anisotropy need not decrease monotonically during inflation: anisotropic stress may continuously source it and \(\sigma/H\) may even grow. What inflation guarantees is that its magnitude remains parametrically small during slow-roll inflation.

The generalized theorem is model independent: it does not predict anisotropic
hair, but allows the possibility that a persistent anisotropic stress may
sustain, or even slowly amplify, a small shear during inflation, subject to the
slow-roll bound above. This possibility is realized, for example, in
vector-driven anisotropic inflation and Higgsed axion--$\mathrm{SU}(2)$
inflation \cite{WatanabeKannoSoda2009,Adshead:2018emn}. Anisotropic stress alone,
however, does not guarantee persistent hair; it must continuously source the
shear. In axion--$\mathrm{SU}(2)$ inflation, for instance, the anisotropic
deformations of the gauge-field configuration decay and the system approaches
the isotropic attractor
\cite{Maleknejad:2011jr,Maleknejad:2013npa,Wolfson:2020fqz,Wolfson:2021fya}.

\section{Conclusions and Outlook}
\label{sec:conclusions-and-outlook}

De~Sitter space has a special place in gravitational physics: it is
simple enough to be treated analytically, yet rich enough to exhibit
cosmological horizons, thermal behavior, observer dependence, and quantum
particle production. It therefore provides a natural meeting point between
geometry, group theory, quantum field theory, and cosmology.

We began in \cref{sec:dS-symmetry} by motivating de~Sitter spacetime as the maximally symmetric geometry
underlying idealized periods of accelerated expansion. We then constructed it
as a hyperboloid embedded in 5D Minkowski space, identified
$\mathrm{SO}(1,4)$ as its isometry group, and showed how de~Sitter symmetry
contracts to Poincar\'e symmetry in the flat-space limit. Then, in \cref{sec:penrose}, the Euclidean
continuation and different Lorentzian coordinate systems were used to
distinguish global properties from patch-dependent ones and to reveal the
causal structure summarized by the Penrose diagram. We next turned to quantum
fields in \cref{sec:dS-qft-casimirs}, organizing one-particle states through the Casimir operators and
unitary representations of the de~Sitter group. Particular attention was
given to the allowed mass ranges, the spin-two sector and the Higuchi bound,
and the relation between bulk masses and late-time conformal weights. Finally, in \cref{sec:particle-production-dS}, we introduced Bogoliubov transformations and particle creation, examined particle production by cosmic expansion, discussed the quantum origin of cosmological perturbations in de~Sitter, and concluded with the cosmic no-hair conjecture and the fate of anisotropies.

The central lesson is that geometry, representation theory, and quantum
dynamics are inseparable in de~Sitter space. The causal structure determines
which regions and observables are accessible to a given observer, the
de~Sitter group organizes the possible field content, and cosmic expansion
shapes the evolution of quantum fields, their late-time behavior, and the
notion of particles.

\subsection{Future Prospects and Open Problems}

In these lecture notes, we have taken only a brief, surface-level tour of
what is currently understood about de Sitter spacetime and quantum fields
propagating on it. Despite decades of progress since de Sitter space first
entered cosmology, the subject remains far from closed. Quantum physics
in de Sitter spacetime continues to be an active area of research, with
several fundamental questions still unresolved. Some of the central open
problems are the following.

\begin{enumerate}

\item \textbf{Observables in de~Sitter quantum gravity.}
In the absence of the usual asymptotic scattering regions, what should
replace the flat-space $S$-matrix? A central challenge is to identify
genuinely gauge-invariant and operational observables and to understand
the relation among global, static-patch, and future-boundary descriptions.
Worldline correlators \cite{Anninos2012StaticPatchSolipsism},
observer-dressed operator algebras \cite{Chandrasekaran2023Observables},
and formulations with finite timelike boundaries
\cite{Anninos2024Observatories} offer complementary approaches.
On global slices, the construction of physical states must also account
for the linearization-stability constraints
\cite{Higuchi1991LinearizationII}.

\item \textbf{Infrared dynamics and semiclassical stability.}
Infrared growth does not by itself imply an instability. Late-time logarithms and their resummation are well understood in several settings \cite{Anninos:2014lwa,Gorbenko:2019rza,Cohen:2020php}, while suitably defined observables can be insensitive to long-wavelength coordinate modes \cite{Senatore:2012nq}. These results should be distinguished from proposals for secular gravitational screening \cite{TsamisWoodard1996}. Whether quantum backreaction can produce a genuine, gauge-invariant departure from de~Sitter expansion remains an open question.

\item \textbf{De~Sitter space and quantum gravity.}
A central open question is whether de~Sitter spacetime admits a controlled
ultraviolet completion \cite{Danielsson:2018ztv,Agmon:2022thq}. Can the
landscape and swampland program \cite{Vafa:2005ui,Ooguri:2006in,
Ooguri:2018wrx}, together with the Weak Gravity Conjecture
\cite{Harlow:2022gzl}, provide decisive criteria? Understanding the resulting
constraints from charged de~Sitter black holes, axion-like fields, and vacuum
decay remains an important step toward answering this question
\cite{Montero:2021otb,Guidetti:2022xct,Kaloper:2022jpv,Kaloper:2023xfl}.

\item \textbf{Unitarity and interacting spinning fields.}
How are representation-theoretic mass bounds, including the Higuchi bound,
modified when exact de~Sitter symmetry is weakly broken? Can partially
massless or higher-spin fields admit consistent interactions, and what
observable signatures would distinguish them?
For the representation-theoretic framework underlying these
questions, including mixed-symmetry and fermionic fields, see
\cite{Hinterbichler2026Representations}.

\item \textbf{Cosmological correlators beyond tree level.}
At tree level, symmetry, locality, analyticity, and unitarity strongly
constrain cosmological correlators and encode the particle content
of inflation \cite{Arkani-Hamed:2015bza,Baumann:2022jpr}.
Important progress also extends beyond tree level, including
loop-resummed four-point functions and non-perturbative unitarity
constraints on late-time correlators
\cite{DiPietro:2021sjt,Hogervorst2023Bootstrap}.
Further developing practical methods for computing and constraining
wavefunction coefficients and in-in correlators, particularly beyond
exact de~Sitter symmetry, remains an important direction.

\item \textbf{Horizon entropy and de~Sitter holography.}
What microscopic degrees of freedom account for the Gibbons--Hawking
entropy? Is there a precise holographic formulation of de~Sitter quantum
gravity, and how should future-boundary and static-patch descriptions
be related?
One-loop bulk and edge contributions provide quantitative probes
of horizon entropy \cite{Anninos2022HorizonEntropy}. Concrete proposals
for counting de~Sitter microstates include $T\bar T+\Lambda_2$
constructions in three dimensions \cite{Coleman2022Microstates}
and a solvable sector of a related deformation in four dimensions
\cite{SilversteinTorroba2025}. Establishing the scope of these
constructions and their relation to local bulk physics remains
an important direction.

\item \textbf{Spectral representations and non-perturbative de~Sitter QFT.}
The de~Sitter K\"all\'en--Lehmann representation provides a non-perturbative
spectral decomposition of two-point functions into unitary representations
of $SO(1,d+1)$ \cite{Loparco:2023KL,DiPietro:2021sjt}. Open questions include
the role of complementary and exceptional sectors, extensions to gravity,
the convergence of boundary expansions, and generalizations beyond
two-point functions and exact de~Sitter space.
Exactly solvable models offer valuable laboratories: the Schwinger
model with a massless charged fermion on $dS_2$ admits non-perturbative,
gauge-invariant correlators and an explicit resummation of late-time
logarithms \cite{Anninos2024Schwinger}.

\item \textbf{The wave function of the Universe.}
The wave function of the Universe encodes a quantum state of geometry
and matter and, in a semiclassical description, provides a framework for
computing late-time correlators of primordial fluctuations.
Its non-perturbative definition, physical boundary conditions,
and the imprint of locality, unitarity, analyticity, and causality remain
central open problems.
The Hartle--Hawking no-boundary proposal
\cite{HartleHawking1983} provides a concrete framework for addressing
these issues through a sum over regular geometries with no initial
boundary.

\item \textbf{Euclidean de~Sitter, analytic continuation, and quantum gravity.}
Euclidean de~Sitter space provides a powerful route to the Bunch--Davies
state and cosmological correlators, yet its role in quantum gravity
remains subtle.
Open questions concern the relevant saddles and contours, the definition
of a Lorentzian state, and the emergence of causal semiclassical spacetime.
Calculations that explicitly include an observer further probe
how the phases and signs of Euclidean gravitational determinants enter
a state-counting interpretation
\cite{Maldacena2024RealObservers,Chen2026DensityOfStates}.
Their extension beyond the settings studied remains an important
direction.

\item \textbf{Dark energy and the late-time approach to de~Sitter.}
The observed cosmic acceleration raises the question of whether the
late-time Universe is described by a cosmological constant or by a
dynamical dark-energy sector. The enormous hierarchy between the observed
vacuum energy and its natural quantum-field-theory scale lies at the heart
of the cosmological-constant problem \cite{Weinberg:1988cp}.
Dynamical alternatives include slowly evolving scalar fields, such as
quintessence \cite{Ratra:1987rm}, while a broad class of dark-energy and
modified-gravity scenarios can be organized systematically within an
effective-field-theory framework \cite{Gubitosi:2012hu,Lewandowski:2016yce}.
More microscopic models can realize distinct late-time dynamics. In quintessence, the required hierarchy appears as an extremely light scalar mass, whereas in non-Abelian gauge-field scenarios such as Gaugessence it is shifted to a very small gauge coupling \cite{Mehrabi:2015hva}. Establishing radiative stability, freedom from instabilities, and observational distinction from $\Lambda$ remains an important challenge.

\end{enumerate}

These questions lie at the heart of quantum physics in de~Sitter space,
where symmetry provides a uniquely controlled setting to connect
representation theory, effective field theory, quantum gravity, and
cosmological observables.

\subsection{Guide to Further Reading}

The following selective references provide useful entry points into the subjects discussed in these notes.

\begin{itemize}[leftmargin=1em]

\item \textbf{Geometry and General Perspective:}

S.~W.~Hawking and G.~F.~R.~Ellis,
\emph{The Large Scale Structure of Space-Time},
Cambridge University Press (1973) \cite{Hawking:1973uf}

M.~Spradlin, A.~Strominger and A.~Volovich,
\emph{Les Houches Lectures on de Sitter Space}, Les Houches Lect. Notes 76 \cite{Spradlin2001}

D.~Anninos,
\emph{De Sitter Musings},
Int.\ J.\ Mod.\ Phys.\ A \textbf{27}, 1230013 (2012)
\cite{Anninos:2012qw}.

D.~A.~Galante, \emph{Modave Lectures on de Sitter Space \& Holography}, PoS Modave 2022 (2023) 003 \cite{Galante2023Modave}.

B.~ Mühlmann
\emph{Notes on de Sitter space},
arXiv:2609.19454
\cite{Muhlmann2026Notes}.

\item \textbf{Quantum Field Theory in Curved Spacetime:}

N.~D.~Birrell and P.~C.~W.~Davies,
\emph{Quantum Fields in Curved Space},
Cambridge University Press (1982) \cite{BirrellDavies1982}

R.~M.~Wald,
\emph{Quantum Field Theory in Curved Spacetime and Black Hole Thermodynamics},
University of Chicago Press (1994) \cite{Wald:1995yp}

L.~Parker and D.~Toms,
\emph{Quantum Field Theory in Curved Spacetime: Quantized Fields and Gravity},
Cambridge University Press (2009) \cite{Parker:2009uva}.

\item \textbf{Representations and Spinning Fields:}

A.~Higuchi,
\emph{Forbidden Mass Range for Spin-2 Field Theory in de Sitter Spacetime},
Nucl.\ Phys.\ B \textbf{282}, 397 (1987) \cite{Higuchi1987}

S.~Deser and A.~Waldron,
\emph{Partial Masslessness of Higher Spins in (A)dS},
Nucl.\ Phys.\ B \textbf{607}, 577 (2001) \cite{Deser:2001pe}

G.~\c{S}eng\"or,
\emph{Particles of a de Sitter Universe},
Universe \textbf{9}, 59 \cite{Sengor:2022kji}.

Z.~Sun, \emph{A Note on the Representations of $\mathrm{SO}(1,d+1)$}, Rev. Math. Phys. 37 (2025) 2430007 \cite{Sun2025}.

\item \textbf{Inflation, Theory and Observations:}

D.~Baumann,
\emph{Primordial Cosmology},
PoS \textbf{TASI2017}, 009 (2018),
arXiv:1807.03098
\cite{Baumann:2018muz}

A.~Maleknejad, M.~M.~Sheikh-Jabbari and J.~Soda,
\emph{Gauge Fields and Inflation},
Phys.\ Rept.\ \textbf{528}, 161--261 (2013)
\cite{Maleknejad:2012fw}.

E.~Komatsu,
\emph{New Physics from the Polarized Light of the Cosmic Microwave Background},
Nature Rev.\ Phys.\ \textbf{4}, 452--469 (2022)
\cite{Komatsu:2022nvu}.

A.~Maleknejad,
\emph{When Geometry Radiates Review: Gravitational Waves in Theory,
Cosmology, and Observation},
arXiv:2512.21328 
\cite{Maleknejad:2025clz}.

\item \textbf{Dark Energy, Theory and Observations:}

 S.~Weinberg, 
\textit{The Cosmological Constant Problem},
Rev.\ Mod.\ Phys.\ \textbf{61} (1989) 1 \cite{Weinberg:1988cp}.

 J.~Frieman, M.~Turner and D.~Huterer,
\textit{Dark Energy and the Accelerating Universe},
Ann.\ Rev.\ Astron.\ Astrophys.\ \textbf{46} (2008) 385 \cite{Frieman:2008sn}.

 T.-N.~Li, G.-H.~Du, H.~Wang, Y.-H.~Li, J.-F.~Zhang and X.~Zhang,
\textit{Dark Energy in the DESI Era: A Brief Review of Evidence,
Beyond-$\Lambda$CDM Interpretations, and Tensions},
arXiv:2606.21826 \cite{Li:2026desi}.

\end{itemize}

\textbf{Acknowledgements} These lecture notes grew out of a mini-lecture series presented by A.M.\ at DESY in February 2026, during her Beate Naroska Guest Professorship at the Cluster of Excellence ``Quantum Universe'', Universit\"at Hamburg. A.M.\ thanks Thomas Konstandin, Geraldine Servant, Alexander Westphal, and the students and researchers who attended the lectures for their insightful questions and discussions, which helped shape these notes. A.M. is particularly grateful to Dionysios Anninos for years of illuminating discussions on de Sitter spacetime and QFT, as well as for his valuable feedback on the draft. A.M.\ also thanks Kamran Salehi Vaziri for his helpful comments on the draft.  A.M.\ is supported by the Royal Society through a University Research Fellowship (Grant No.~RE22432). The lectures and related work were supported by the Deutsche Forschungsgemeinschaft under Germany's Excellence Strategy (EXC 2121 ``Quantum Universe'' -- 390833306).

\appendix
\crefalias{section}{appendix}

\include{Open-coord}

\include{bianchi}

\bibliographystyle{JHEP}
\bibliography{ref.bib}

\end{document}

%% file: Open-coord.tex
\section{The Open FLRW Slicing of de Sitter Spacetime}\label{sec:open-coord}

For completeness, in this appendix we briefly discuss the open slicing of de Sitter spacetime, obtained by parametrizing the de Sitter hyperboloid as
\begin{align}
X_0
&=
H^{-1}\sinh(Ht)\cosh\chi,
\\
X_i
&=
H^{-1}\sinh(Ht)\sinh\chi\,\hat n_i,
\qquad i=1,2,3,
\\
X_4
&=
H^{-1}\cosh(Ht),
\end{align}
where $\sum_{i=1}^{3}\hat n_i^2=1$. Thus, it is related to a FLRW metric with
\begin{equation}
a(t)=H^{-1}\sinh(Ht),
\qquad
k=-1.
\end{equation}

For the expanding open patch, one takes $t>0$. Each constant-$t$ hypersurface is then a non-compact copy of $\mathbb{H}^3$, with intrinsic scalar curvature
\begin{equation}
{}^{(3)}R
=
-\frac{6}{a^2(t)}
=
-\frac{6H^2}{\sinh^2(Ht)},
\end{equation}
and the corresponding FLRW Hubble parameter is
\begin{equation}
\frac{\dot a}{a}
=
H\coth(Ht).
\end{equation}
At late times,
\begin{equation}
a(t)
\simeq
\frac{1}{2H}e^{Ht},
\end{equation}
so the expansion asymptotically approaches the exponential behavior of the flat slicing.

As $t\to0^+$, the scale factor vanishes. This is not a curvature singularity of the four-dimensional spacetime: the de Sitter curvature invariants remain finite and constant. Rather, the open foliation degenerates at a coordinate boundary, analogous to the origin of Milne coordinates in Minkowski spacetime (see the right panel of \cref{fig:flat-open-close}). Unlike the global coordinates, the open coordinates cover only a restricted region of the de Sitter hyperboloid. For the expanding branch,
\begin{equation}
X_4\geq H^{-1},
\qquad
X_0>0.
\end{equation}
The branch $t<0$ describes the corresponding contracting region, while an analogous open chart may be introduced for the antipodal region $X_4\leq-H^{-1}$. Open coordinates are particularly natural in problems involving vacuum decay and bubble nucleation. In particular, the interior of a Coleman--De Luccia bubble admits homogeneous and isotropic spatial slices of negative curvature, for which the open FLRW description is naturally adapted.

%% file: bianchi.tex

\section{The Bianchi Family of Homogeneous Cosmologies}
\label{sec:bianchi}

Spatial homogeneity does not require the metric components to be independent of
the spatial coordinates in an arbitrary coordinate basis. Rather, it requires
the existence of a group of spatial isometries acting transitively on each
constant-time hypersurface. The natural framework for describing such
geometries is provided by the Bianchi family of homogeneous cosmological
spacetimes. Here, we briefly introduce their geometrical construction and their relation to homogeneous
anisotropies.

A Bianchi spacetime admits a foliation by three-dimensional homogeneous
hypersurfaces,
\begin{equation}
    \mathcal{M}=\mathbb{R}\times\Sigma_t ,
\end{equation}
on each of which a three-dimensional Lie group acts transitively. Let
$\{\boldsymbol e_a\}$ be an invariant spatial frame and
$\{\boldsymbol\omega^a\}$ its dual coframe,
\begin{equation}
    [\boldsymbol e_a,\boldsymbol e_b]
    =C^{c}{}_{ab}\boldsymbol e_c,
    \qquad
    \boldsymbol\omega^a(\boldsymbol e_b)=\delta^a{}_b .
\end{equation}
In synchronous time, the most general spatially homogeneous metric may be
written as
\begin{equation}
    \mathrm{d}s^2
    =
    -\mathrm{d}t^2
    +a^2(t)\,
    \bigl[\exp(2\boldsymbol\beta(t))\bigr]_{ab}
    \boldsymbol\omega^a\boldsymbol\omega^b,
    \qquad
    \operatorname{tr}\boldsymbol\beta=0,
    \label{eq:general-bianchi-metric}
\end{equation}
where $a(t)$ describes the isotropic expansion, while the symmetric
traceless matrix $\beta_{ab}(t)$ parametrizes the anisotropy. The Bianchi
type is determined by the structure constants $C^{c}{}_{ab}$ and is
preserved under time evolution. As summarized in
Table~\ref{tab:bianchi-families}, only a subset of the Bianchi families
contains exact FLRW geometries as isotropic limits: the spatially flat
FLRW geometry belongs to types~I and~$\mathrm{VII}_{0}$, the open
geometry to types~V and~$\mathrm{VII}_{h}$, and the closed geometry to
type~IX \cite{Bianchi1898,Stephani2003}.

Exact de~Sitter spacetime can therefore arise within these five Bianchi
families, depending on the choice of spatial slicing. Writing
$H^2=\Lambda/3$, its FLRW scale factor takes the form
\begin{equation}
    a_{\mathrm{dS}}(t)
    =
    \begin{cases}
        a_0 e^{Ht},
        & K=0,
        \qquad \text{types I and $\mathrm{VII}_{0}$},
        \\[1mm]
        H^{-1}\sinh(Ht),
        & K=-1,
        \qquad \text{types V and $\mathrm{VII}_{h}$},
        \\[1mm]
        H^{-1}\cosh(Ht),
        & K=+1,
        \qquad \text{type IX}.
    \end{cases}
    \label{eq:de-sitter-bianchi-slicings}
\end{equation}
These represent different foliations of the same maximally symmetric
de~Sitter spacetime (see \cref{subsec:dS-slicings}). Their existence as exact
isotropic members should be distinguished from dynamical isotropization:
Bianchi geometries that do not contain an exact FLRW limit may nevertheless
approach de~Sitter asymptotically under the conditions of the cosmic no-hair
theorem. In the following, we briefly review Bianchi types~$\mathrm{I}$ and
$\mathrm{VII}_{0}$, both of which contain the spatially flat FLRW geometry as
an isotropic limit.

\begin{table}[h]
\centering
\small
\renewcommand{\arraystretch}{1.2}
\begin{tabular}{c p{6cm} c}
\hline
\textbf{Bianchi type}
&
\textbf{Spatial geometry}
&
\textbf{FLRW limit}
\\
\hline

$\mathrm{I}$
& Flat, translation-invariant
& $K=0$
\\

$\mathrm{II}$
& Twisted homogeneous geometry
& None
\\

$\mathrm{III}$
& $\mathbb{H}^{2}\times\mathbb{R}$--type geometry
& None
\\

$\mathrm{IV}$
& Homogeneous solvable geometry
& None
\\

$\mathrm{V}$
& Hyperbolic geometry
& $K<0$
\\

$\mathrm{VI}$
& Anisotropic hyperbolic-type geometry
& None
\\

$\mathrm{VII}$
& Homogeneous spiralling geometry
& $\begin{array}{c}
K=0\quad (\mathrm{VII}_{0})\\
K<0\quad (\mathrm{VII}_{h})
\end{array}$
\\

$\mathrm{VIII}$
& $\widetilde{\mathrm{SL}}(2,\mathbb{R})$ geometry
& None
\\

$\mathrm{IX}$
& Spherical, compact geometry
& $K>0$
\\

\hline
\end{tabular}
\caption{
The nine Bianchi families and their isotropic FLRW limits. Only
types~$\mathrm{I}$, $\mathrm{V}$, $\mathrm{VII}$, and $\mathrm{IX}$
contain exact FLRW geometries: types~$\mathrm{I}$ and
$\mathrm{VII}_{0}$ contain the spatially flat case,
types~$\mathrm{V}$ and $\mathrm{VII}_{h}$ the open case, and
type~$\mathrm{IX}$ the closed case. Accordingly, these families also
contain the flat, open, and closed slicings of exact de~Sitter spacetime,
respectively.
}
\label{tab:bianchi-families}
\end{table}

\paragraph{Bianchi type I.}
The type-I algebra is Abelian,
\begin{equation}
    [\boldsymbol e_a,\boldsymbol e_b]=0,
\end{equation}
so one may choose $\boldsymbol\omega^1=\mathrm{d}x$, $\boldsymbol\omega^2=\mathrm{d}y$, and
$\boldsymbol\omega^3=\mathrm{d}z$.  When the spatial metric is diagonal, one obtains
\begin{equation}
    \mathrm{d}s^2=-\mathrm{d}t^2
    +a_1^2(t)\mathrm{d}x^2+a_2^2(t)\mathrm{d}y^2+a_3^2(t)\mathrm{d}z^2.
    \label{eq:bianchi-I}
\end{equation}
Thus type~I describes a homogeneous anisotropy whose principal axes have no
spatial twist.

\paragraph{Bianchi type $\mathrm{VII}_{0}$.}
The Euclidean group in two dimensions underlies the type-$\mathrm{VII}_{0}$
algebra.  Keeping an inverse-length scale $\kappa$ explicit, its Killing
generators can be chosen to obey
\begin{subequations}
\begin{align}
    [X_1,X_2]&=0,\\
    [X_1,X_3]&=\kappa X_2,\\
    [X_2,X_3]&=-\kappa X_1,
\end{align}
\end{subequations}
with
\begin{equation}
    X_1=\partial_x,
    \qquad
    X_2=\partial_y,
    \qquad
    X_3=\partial_z
    +\kappa\bigl(x\partial_y-y\partial_x\bigr).
    \label{eq:VII0-killing}
\end{equation}
The third isometry is therefore a translation along the $z$ direction combined
with a rotation in the transverse plane; its integral curves are helices.

A convenient invariant frame is
\begin{subequations}
\begin{align}
    \boldsymbol e_1&=\cos(\kappa z)\,\partial_x
             +\sin(\kappa z)\,\partial_y,\\
    \boldsymbol e_2&=-\sin(\kappa z)\,\partial_x
             +\cos(\kappa z)\,\partial_y,\\
    \boldsymbol e_3&=\partial_z,
\end{align}
\end{subequations}
with dual one-forms
\begin{subequations}
\begin{align}
    \boldsymbol\omega^1&=\cos(\kappa z)\,\mathrm{d}x
                  +\sin(\kappa z)\,\mathrm{d}y,\\
    \boldsymbol\omega^2&=-\sin(\kappa z)\,\mathrm{d}x
                  +\cos(\kappa z)\,\mathrm{d}y,\\
    \boldsymbol\omega^3&=\mathrm{d}z.
\end{align}
\end{subequations}
Although these basis elements rotate with $z$, their Lie transport along the
Killing fields vanishes.  Consequently, a metric whose components depend only
on time in this frame is spatially homogeneous.